\documentclass[conference,10pt]{IEEEtran}
\usepackage{amsmath,amssymb}
\usepackage{graphicx}
\usepackage{booktabs}
\usepackage{url}
\makeatletter
\g@addto@macro{\UrlBreaks}{\do\-}
\makeatother
\usepackage{xcolor}
\usepackage{placeins}
\usepackage{stfloats}
\usepackage{comment}

\begin{document}
\title{Closing the Prediction Gap: Completing Machine Shape So That
Predicted Time, Power, Energy, and Mapping Match What Real Hardware
Does}

\author{
\IEEEauthorblockN{Lenore M. Mullin}
\IEEEauthorblockA{Professor Emerita, College of Nanotechnology, Science and Engineering\\
University at Albany, SUNY, Albany, NY\\
lmullin@albany.edu}
\and
\IEEEauthorblockN{Ga\'etan Hains}
\IEEEauthorblockA{LACL, Universit\'e Paris-Est Cr\'eteil\\
Cr\'eteil, France\\
gaetan.hains@u-pec.fr}
}

\maketitle
\pagestyle{plain}

\begin{abstract}
A companion paper derives a weighted, communication-free partition for
heterogeneous, multi-institution device ensembles from each device's measured
shape, $\rho_{\mathrm{machine}}(d)$, and validates its predictions on real
hardware. Its experiments, however, reveal that
$\rho_{\mathrm{machine}}$ is incomplete: fp32-to-fp16 speedup and power shifts
are not fully indexed, and memory-capacity predictions depend on an assumed
rather than measured reservation overhead.

This paper extends the theory by indexing five previously unmeasured fields in
the same $\psi$-selection vocabulary: memory capacity
$M_i^{\mathrm{cap}}$, power draw $P_i$, execution-unit capability $X_d$,
compiler-directive configuration $C_d$, and memory-reservation overhead
$R_d$. We introduce $\mathrm{reached}(d)$, a Boolean indicating whether
compiled code actually dispatches to specialized hardware. This distinction
explains behavior such as NVIDIA's fp16 power shift, which cannot be predicted
from hardware availability alone. We restate the companion paper's time and
energy models in terms of $\mathrm{reached}(d)$, rather than merely the
presence of specialized execution units.

Because MoA fixes a computation independently of its target, we define a
two-part admissibility criterion for compiler directives: a directive must
select a real hardware quantity without discretionary reordering, and it must
be deterministic rather than compiler-overridable. Across Intel, AMD, NVIDIA,
OpenMP, OpenACC, and Open MPI, this leaves a small admissible class:
cache-placement controls, deterministic vector or tile widths, and occupancy
or placement caps. Heuristic mechanisms, including auto-vectorization and
autotuning enable flags, are excluded and represented by a residual
$\epsilon_d^{(k)}$.

We validate the extended model on five production devices: NVIDIA A100 and
H100, AMD MI100, NVIDIA V100, and Intel Max 1550. On each device, we directly
measure or independently establish all five fields. Results show that
specialized-hardware reachability can be real but partial, occupancy limits
must be established per device and admissible realization, and
memory-reservation overhead ranges from approximately $0.004$ to $0.030$. We
also measure a negative residual of $-9.8\%$ for Triton's excluded autotuner,
demonstrating that an unvalidated heuristic can have an effect opposite to a
presumed penalty. Together, these experiments close all five fields for every
device in the companion paper's ensemble and convert previously assumed or
unindexed behavior into measured, target-specific quantities.
\end{abstract}
\section{Introduction}
A prediction that is only checked where it happens to already be
right is not a validated theory; it is a coincidence reported as one.
The companion paper to this one~\cite{mullin2026hetero} does not make
that mistake: every one of its predicted numbers -- a finish time, a
memory ceiling, an energy figure -- is checked against real,
measured hardware, and two of those checks come back wrong. Both
failures are diagnosed precisely, in closed form, rather than patched
around, and both diagnoses point to the same conclusion stated only
in that paper's own future-work section: $\rho_{\text{machine}}(d)$,
the tuple every prediction in that paper's theory is built from, is
missing fields that its own real experiments already show are
load-bearing.

This paper's job is narrow and stated precisely: close those fields,
giving the missing quantities the same formal status every
already-indexed field of $\rho_{\text{machine}}$ has -- a name, a
$\psi$-selectable index position, and a stated role in a closed-form
prediction -- reasoning throughout from the companion paper's own
real numbers. A theory that names a gap in prose but leaves the
corresponding equation silently incomplete has not closed anything;
this paper's contribution is closing it in the equations themselves,
and, for one device, checking that closure against a first real
measurement (Section~\ref{sec:validation-a100}) rather than leaving
the extended theory entirely unvalidated.

\begin{figure*}[t]
\centering
\includegraphics[width=0.85\linewidth]{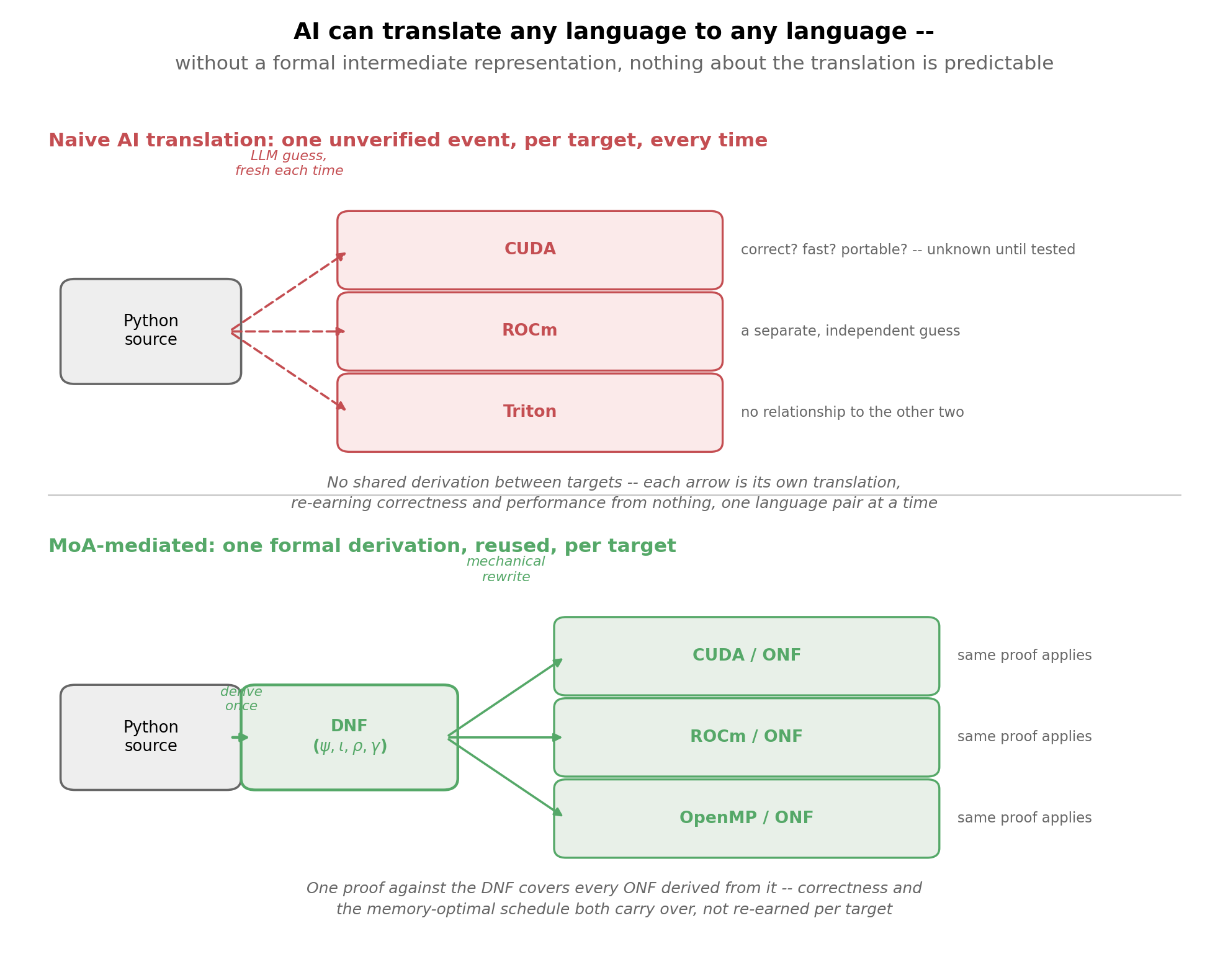}
\caption{Why a formal intermediate representation, not just a
capable translator: naive AI translation (top) produces a fresh,
unverified guess per target language, with no relationship between
one translation and the next. MoA-mediated translation (bottom)
derives one formal specification once (the DNF) and mechanically
produces each target's realization (its ONF) from it, so correctness
and the memory-optimal schedule both carry over rather than being
re-earned per target -- the argument developed and tested on real
hardware in~\cite{mullin_ipdps2027}.}
\label{fig:ai-vs-moa}
\end{figure*}

Why any of this matters, stated once at the outset for a reader who
has not yet been asked to care about a specific index position in a
specific tuple: large language models can already translate source
code from any language to any other, fluently, on request. What they
cannot do is guarantee that the result is correct, memory-optimal, or
predictable on the hardware it actually runs on -- each translation is
its own unverified event, earning nothing from the last one
(Figure~\ref{fig:ai-vs-moa}). A formal intermediate representation is
what turns fluent translation into trustworthy translation: derive a
computation's meaning once, independent of target (MoA's DNF), and
mechanically produce a verified, memory-optimal realization per
target (its ONF) -- one proof reused, not re-earned per language,
per vendor, per device. Prior work in this research program argues
this case directly and tests it against production kernels on real
hardware~\cite{mullin_ipdps2027}. $\rho_{\text{machine}}(d)$ is the
other half of that same argument: a formal derivation is only as
trustworthy as the machine-shape facts it is derived against, and
this paper's job, narrow as it is, is making sure none of those facts
are still missing.

\subsection{MoA in Brief: From Array Data to Machine Shape}
\label{sec:moa-brief}
This subsection is for a reader meeting the Mathematics of Arrays
(MoA) for the first time; a reader already fluent in it can skip
directly to Section~\ref{sec:five-fields}.

MoA~\cite{mullin1988} is a formal array calculus: every array is a
pair (shape, values), every array operation is a composition of a
small, fixed set of algebraic operators -- $\psi$ (index/select),
$\iota$ (generate an index set), $\rho$ (shape), $\gamma$
(index-substitution, i.e.\ partition or reshape), and $\mathrm{rav}$
(ravel to one dimension) -- and a whole computation is first written
once as a Denotational Normal Form (DNF): a single, target-independent
composition of these operators that states \emph{what} is computed,
with no loop order, tiling, or device assumed. A mechanical rewrite
step, justified by the algebra's own identities rather than by
hand-tuning, turns that DNF into an Operational Normal Form (ONF) --
a loop nest, or a target-offload kernel -- for a specific device. The
goal this whole research program serves is stated plainly: derive
correct, portable, and \emph{predictable} low-level code from one
formal specification, instead of hand-writing (and separately
debugging) one implementation per vendor, and instead of leaving a
kernel's performance to be discovered empirically after the fact.
Every kernel measured in this paper and its companion is ONF code
produced this way, not hand-tuned by trial and error.

The companion paper's own contribution, which this paper takes as
given, is to apply this same DNF/ONF/$\gamma$ machinery once more, one
level up: instead of partitioning an array across the lanes of a
single device, it partitions an array across the devices of a
heterogeneous, multi-institution ensemble, with each device's share
sized by that device's own measured shape, $\rho_{\text{machine}}(d)$
-- exactly as Figure~\ref{fig:baby-machines} shows for a toy instance
and Figure~\ref{fig:array-machine-mapping} shows for a real, measured
one. $\rho_{\text{machine}}(d)$ is, formally, nothing more exotic than
another MoA array: a tuple, indexed by $\psi$, whose entries happen to
be hardware facts (compute-unit count, memory ceiling, measured time)
rather than tensor values. Table~\ref{tab:indexed-fields} below
restates its five fields as the companion paper already indexes them;
this paper's entire contribution is a \emph{dimension lift} on that
same array -- adding five more index positions to it, using that same
$\psi$-selection notation, rather than introducing any new formalism
-- so that quantities the companion paper's own real hardware
experiments already showed mattered -- memory capacity, power draw,
which execution units a compiled kernel actually reaches, which
compiler directives are deterministic enough to trust, and how much
memory is reserved rather than usable -- have a formal place to live,
instead of being silently folded into an existing field or left as an
unindexed assumption. Formulating that lift precisely, deriving the
predictions it feeds algebraically, and validating both against real
hardware is this paper's entire arc, restated once here: formulate the
five fields as index positions (Section~\ref{sec:five-fields}), derive
what they complete in the companion paper's own equations
(Sections~\ref{sec:five-fields}--\ref{sec:reached}), state precisely
what stays open to compiler or runtime optimization versus what is
now shape-fixed (Section~\ref{sec:cd-survey} onward), and predict,
then validate, every one of those five fields on real hardware for
every device in the companion paper's ensemble
(Sections~\ref{sec:validation-a100}--\ref{sec:validation-pvc}).

\subsection{The two real findings this paper closes}
The companion paper's Section on lower-precision realization
(its Section~IX-G) measures fp32-to-fp16 speedup on all three vendors
in its ensemble at the identical reference problem size: $14.7\times$
on NVIDIA (via a Triton kernel whose fused \texttt{tl.dot} reaches the
device's tensor cores directly), $1.05\times$ on Intel, and no
speedup at all, within measurement noise, on AMD (both realized as
portable OpenMP target-offload code with no path to either vendor's
specialized low-precision hardware). Its energy section (Section~IX-H)
measures the same three devices' power draw at fp32 versus fp16 and
finds the identical split restated on a second, independent axis:
NVIDIA's power draw rises $36.3\%$ between precisions, while Intel's
and AMD's shift by less than a percent each -- not because one vendor's
silicon is more capable in kind, but because one realization actually
reached specialized hardware built for that precision and the other
two did not.

Nothing in $\rho_{\text{machine}}(d)$ as defined -- five fields,
compute units, warp width, shared memory, measured time, memory
ceiling -- names this. The companion paper says so explicitly: it
names the missing field $X_d$, execution-unit-type, and immediately
qualifies it as conflating two different questions that need
separating before either is trusted -- whether the hardware exists at
all, and whether a specific kernel's compiled realization actually
dispatches to it. That second question is a property of how the
kernel was compiled, not of the device, and the companion paper names
a second missing field for it, $C_d$, compiler directives, without
populating it. This paper populates both, and states the interaction
between them as the single quantity the energy and time predictions
actually depend on.

The companion paper's memory-ceiling prediction (its Equation~39) has
a second, quieter gap. Its overhead term -- the fraction of a
device's memory capacity reserved for driver and runtime footprint,
not available to hold the key/value cache -- was initially assumed at
a conservative $15\%$, found wrong (a unit error, decimal gigabytes
read where the device reports binary gibibytes, was the larger
effect), and re-evaluated at $0\%$ once that unit error was corrected.
Zero is what one corrected evaluation on one device happened to need;
it is not a measured property of memory reservation in general, and
the companion paper does not claim it is. This paper names overhead
as a per-device field for the same reason it names $M_i^{\text{cap}}$
itself: a quantity three different real bugs already turned out to
depend on deserves an index position, not an assumed constant reused
silently at every future call site.

\begin{figure*}[t]
\centering
\includegraphics[width=0.78\linewidth]{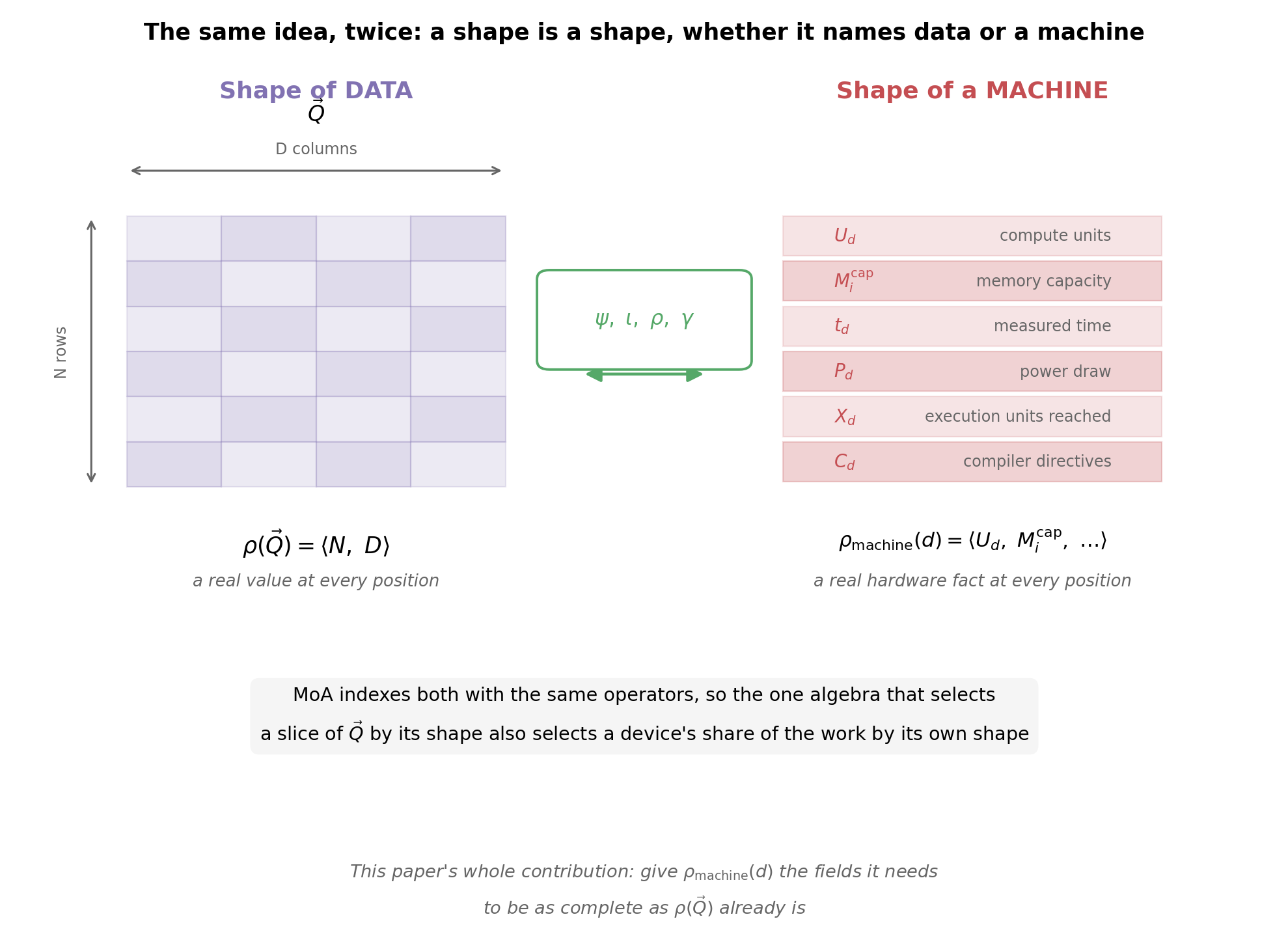}
\caption{The idea this whole paper rests on, stated before any operator
notation: MoA treats the shape of an array, $\rho(\vec{Q})=\langle N,D\rangle$,
and the shape of a machine, $\rho_{\mathrm{machine}}(d)$, as the same
kind of object -- a tuple with a real value at every position, indexed
by the same operators ($\psi,\iota,\rho,\gamma$). This is why deriving
a kernel from an array's shape and deriving a device's share of the
work from that device's own shape is one algebra, not two. This
paper's contribution is exactly the right-hand side: giving
$\rho_{\mathrm{machine}}(d)$ the fields it needs to be as complete as
$\rho(\vec{Q})$ already is.}
\label{fig:shapes-data-machines}
\end{figure*}

\begin{figure*}[t]
\centering
\includegraphics[width=0.72\linewidth]{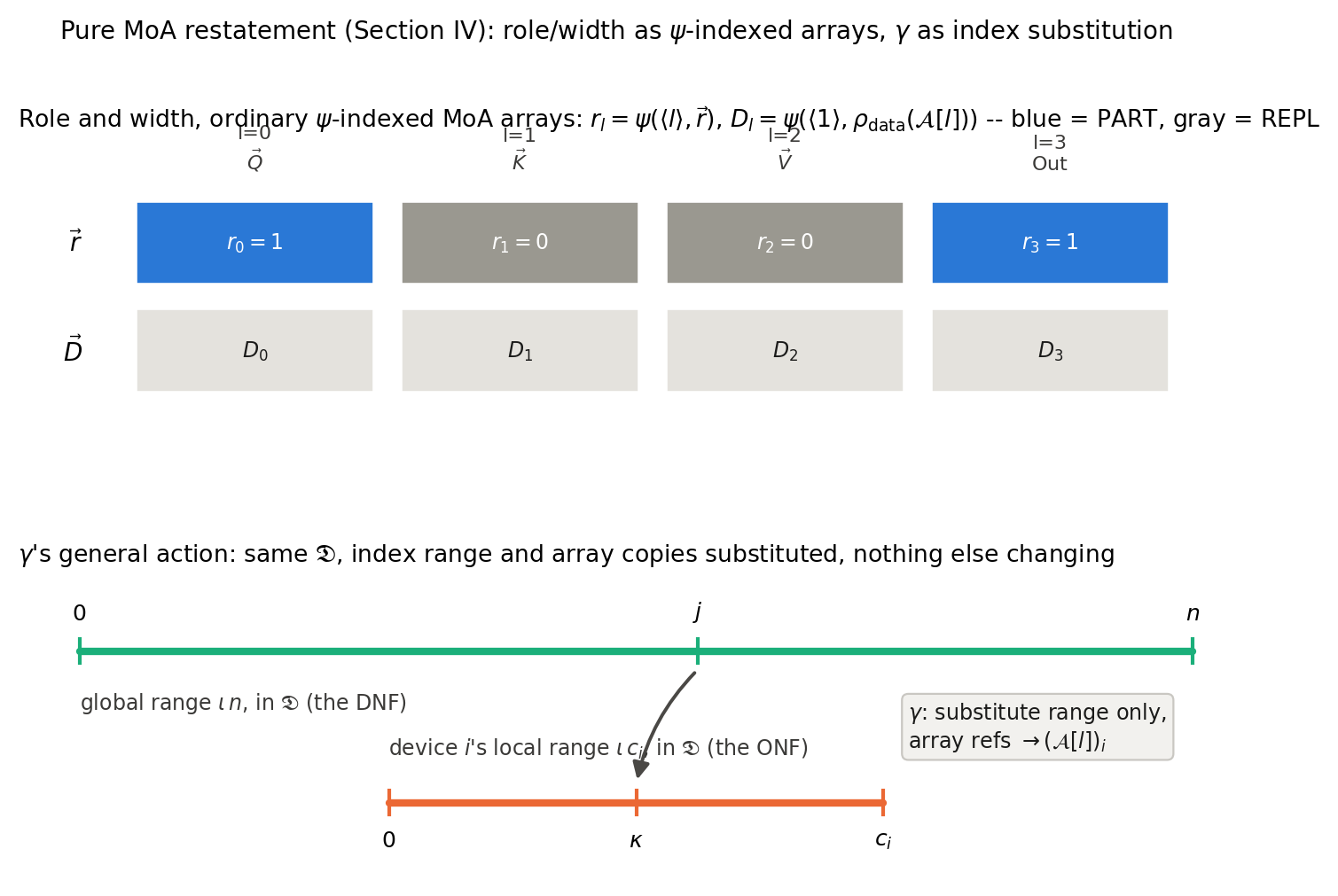}
\caption{The two MoA operators this paper leans on most, made concrete.
Top: an array's \emph{role} ($r_l\in\{0,1\}$, partitioned vs.\
replicated across devices) and \emph{width} ($D_l$) are themselves
just $\psi$-indexed MoA arrays -- role and width are data, not
metadata bolted on separately. Bottom: $\gamma$ is the single
substitution that turns a device-local index $j\in\iota\,c_i$ into
its corresponding position $j+o_i$ in the shared global range $\iota\,n$
-- the same operator, applied once, that later reappears as the
mechanism behind every device's own $(o_i,c_i)$ slice below.}
\label{fig:pure-moa}
\end{figure*}

\begin{figure*}[t]
\centering
\includegraphics[width=0.75\linewidth]{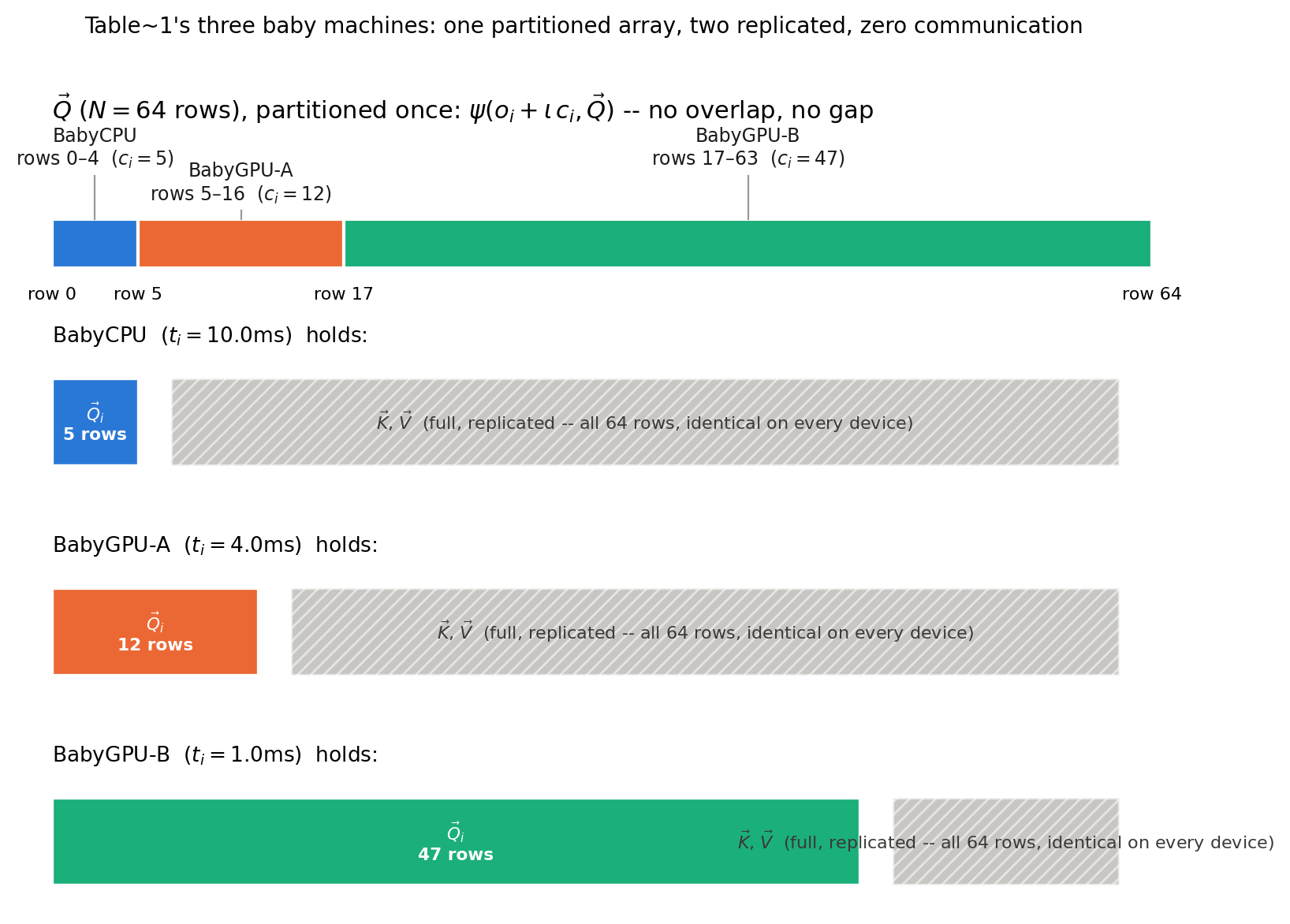}
\caption{A concrete, minimal instance -- a toy $64$-row query array
split across three invented devices, sized inversely to each device's
own measured time $t_{d_i}$ (Equation~26 of the companion paper). Top:
$\vec{Q}$ is cut into three disjoint, unequal $\psi$-selected slices,
one per device, via the offset/count pair $(o_i,c_i)$. Bottom: what
each device actually holds -- its own unique $\vec{Q}$ slice, plus a
full, identical, \emph{replicated} copy of $\vec{K}$ and $\vec{V}$
(hatched), since attention's key/value tensors are read whole by
every device rather than partitioned. This is the same partition
structure Section~\ref{sec:validation-a100} onward measures on real
hardware, at real problem sizes, rather than a toy count.}
\label{fig:baby-machines}
\end{figure*}

\begin{figure*}[t]
\centering
\includegraphics[width=0.85\linewidth]{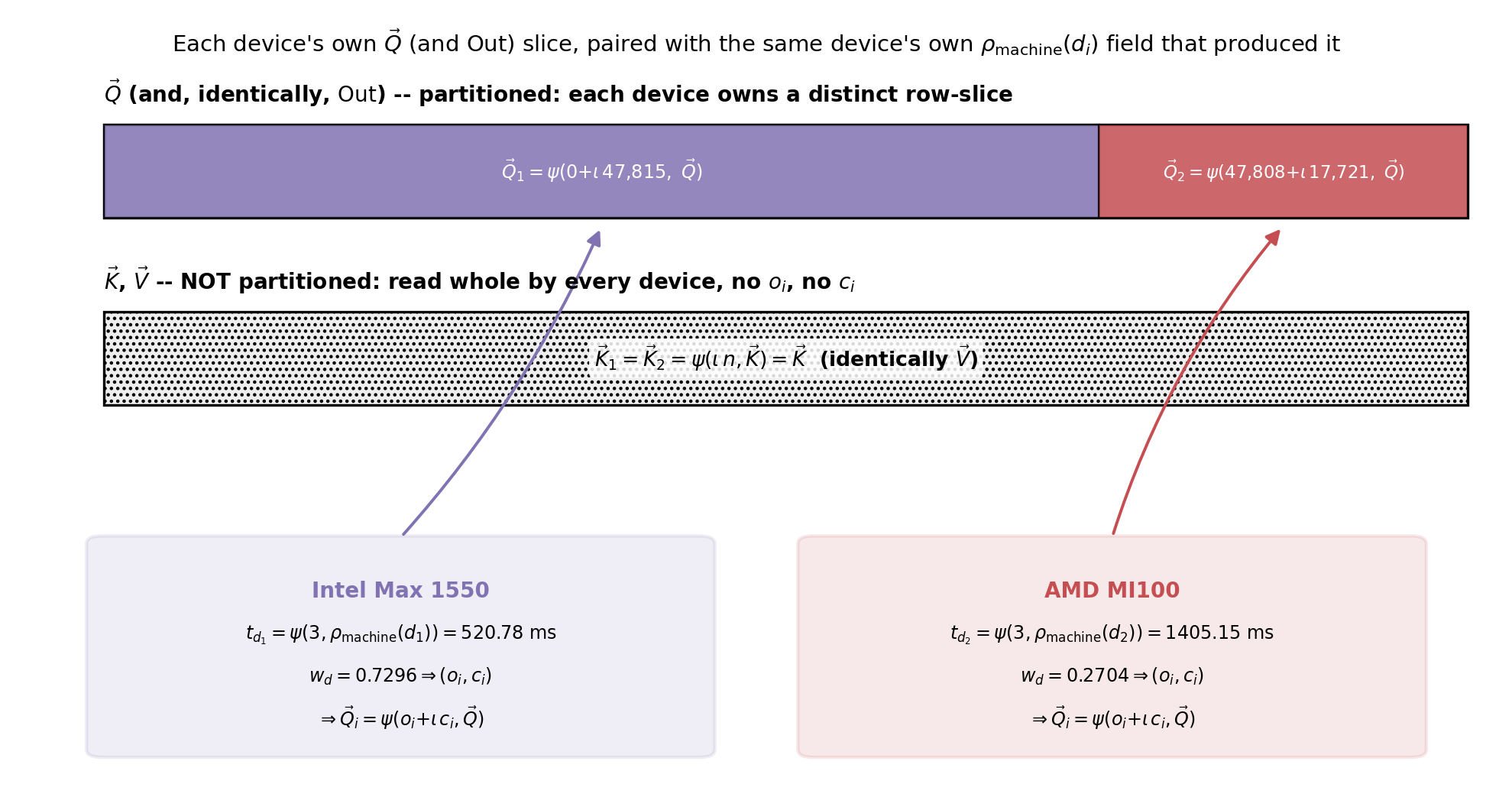}
\caption{Array data selected \emph{by} array machine information, on
a real, validated two-device run (Intel Max 1550 + AMD MI100,
$n=65{,}536$, companion paper). Each device's own row-slice of
$\vec{Q}$ (and $\mathrm{Out}$) is produced from that same device's own
$\psi$-selected $\rho_{\text{machine}}(d_i)$ entry -- concretely,
$t_{d_i}=\psi(\langle3\rangle,\rho_{\text{machine}}(d_i))$ feeds the
weight $w_{d_i}$, which in turn fixes $(o_i,c_i)$ and hence the exact
slice $\vec{Q}_i=\psi(o_i{+}\iota\,c_i,\vec{Q})$ that device computes
on. $\vec{K}$ and $\vec{V}$, by contrast, are identical and unsliced
on both devices: no $\rho_{\text{machine}}(d_i)$ entry participates in
selecting them at all. This is the exact mechanism this paper's five
new fields plug into: each is a new $\rho_{\text{machine}}(d_i)$
entry, selected by the identical $\psi$ notation already driving the
partition above, not a new kind of quantity requiring new notation.}
\label{fig:array-machine-mapping}
\end{figure*}

The four figures below make this concrete in order:
Figure~\ref{fig:shapes-data-machines} states the single idea everything
else in this paper builds on -- that a shape is a shape, whether it
names an array or a device -- before any operator notation is assumed;
Figure~\ref{fig:pure-moa} then restates the operators themselves,
Figure~\ref{fig:baby-machines} shows a toy partition, and
Figure~\ref{fig:array-machine-mapping} shows the same partition on a
real, measured run.

\subsection{What this paper does and does not claim}
This paper's own real content is the formulation: naming five fields
precisely, stating exactly which existing equation each one completes,
and showing algebraically that the resulting predictions reduce to
the companion paper's own equations in the special case where the
new fields take the values that paper's real measurements already
established. Every number quoted from the companion paper is quoted, not
re-derived, and every claim about what a real compiler toolchain
exposes is sourced to that toolchain's own
documentation~\cite{intel_prefetch_gpu_2025,
intel_qopt_prefetch_2024, intel_xe_arch_2025,
rocm_compiler_ref_2024, amd_matrix_cores_gpuopen, triton_config_docs},
not assumed. For one device, NCSA Delta's A100, this paper also runs
a new round of hardware experiments of its own
(Section~\ref{sec:validation-a100}), validating the extended theory
directly rather than only formally, with all five fields measured. A
second experiment, on Delta's AMD MI100
(Section~\ref{sec:validation-mi100}), validates $C_d$'s occupancy-cap
field and the $X_d$/$C_d$ interaction specifically, under this
paper's own admissible \texttt{vector\_size} realization rather than
the companion paper's excluded \texttt{simdlen} one, and is now
followed, in the same section, by direct measurement of the remaining
two fields, $R_d$ and $X_d$, closing all five on this device. Two
further real-hardware experiments, on PSC Bridges-2's H100
(Section~\ref{sec:validation-h100}) and SDSC Expanse's V100
(Section~\ref{sec:validation-v100}), each close all five fields as
well, using a different, vendor-neutral OpenACC realization instead
of Triton -- a test of the method itself, not only of one device or
one kernel family. A fifth real-hardware experiment, on TACC
Stampede3's Intel Max 1550, the one remaining device in the companion
paper's ensemble (Section~\ref{sec:validation-pvc}), closes this
device in full as well: $M_i^{\text{cap}}$ and $C_d$'s occupancy-cap are
measured and pinned on real hardware, $P_d$ reads consistently across
repeated runs, $\mathrm{reached}(d)$ is confirmed directly at
$99.88\%$ busy via Level Zero Sysman engine activity, $X_d=0$
follows from the kernel's own construction, and $R_d\approx0.004$ is
located by the same bracket-and-locate sweep used on the other three
GPUs -- all five devices in the companion paper's ensemble now closed
in full, the next step named precisely in Section~\ref{sec:future}.

\section{Related Work: Why a Formal, Deterministic Account of Machine Shape}
\label{sec:related}
Two mature literatures already predict or optimize GPU kernel
performance; both differ from this paper (and its companion) in the
same structural way, which is worth stating precisely rather than
only claiming.

\textbf{Autotuning compilers search; MoA derives.} Halide~\cite{ragankelley2013halide}
separates an image-processing pipeline's algorithm from its
\emph{schedule} (loop order, tiling, fusion), then finds a good
schedule by search -- originally manual, later autotuned. TVM~\cite{chen2018tvm}
and Ansor~\cite{zheng2020ansor} generalize this to deep-learning
operators: a large schedule space is defined, and a cost model plus
on-device measurement searches it, at real, nontrivial compute and
wall-clock cost, and with no closed-form guarantee that the mechanism
producing a fast schedule on one device transfers to another. This
paper's $C_d$-admissibility criterion (Section~\ref{sec:cd-survey}) makes
the same distinction those systems draw operationally -- some
schedule choices are safe, deterministic layout decisions, others are
heuristics the compiler or autotuner is free to override -- but as a
formal predicate checked once against a mechanism's own documentation
and hardware manual, not as a search that must be re-run per device,
per kernel, and per compiler version. MoA's DNF/ONF derivation
(Section~\ref{sec:moa-brief}) plays the role Halide's schedule and
TVM/Ansor's search space play, except that the admissible
transformations are fixed by the algebra's own identities in advance,
so a specific $C_d$ configuration is a \emph{stated, checkable}
fact about a realization rather than the output of a search whose
result is trusted because it measured fast on one occasion.

\textbf{Performance models bound; $\rho_{\text{machine}}(d)$ predicts,
and is checked.} The Roofline model~\cite{williams2009roofline} plots
a kernel's arithmetic intensity against a device's peak bandwidth and
peak compute to bound achievable performance -- a real, widely used
tool, but a bound, not a specific predicted time, and one that says
nothing about whether a given compiled kernel actually dispatches to
the execution units that peak figure assumes. That is precisely the
gap $\mathrm{reached}(d)$ (Section~\ref{sec:reached}) closes: this
paper's five new fields, together with the companion paper's
original five, do not bound a family of possible kernels, they
predict one specific kernel's time, power, and memory ceiling on one
specific device, and every validation section in this paper
(Sections~\ref{sec:validation-a100}--\ref{sec:validation-v100})
checks that specific prediction against a live, direct hardware
measurement -- correctness, utilization-confirmed dispatch, and
timing/power/memory jointly -- rather than checking only that a
measured point falls under a bound.

\textbf{Hand-tuned kernel libraries achieve speed on one device at a
time; this program states, and tests, whether the same derivation
transfers.} FlashAttention~\cite{dao2022flashattention} is the
clearest example in the attention-kernel space this paper's own
validation targets: an IO-aware, fused kernel, hand-designed around
one specific memory hierarchy (on-chip SRAM versus HBM) to avoid
materializing the full attention matrix, achieving real, substantial
speedups. It is not, and does not claim to be, a portable derivation
-- porting its specific tiling and recomputation strategy to a new
vendor's memory hierarchy is itself new engineering work, repeated
per device family (NVIDIA's own later kernels for AMD and Intel
targets confirm this: they are separate implementations, not one
schedule retargeted). The companion paper's own OpenMP/Triton/OpenACC
realizations, and this paper's extension of $\rho_{\text{machine}}(d)$
to explain why they behave differently per vendor, take the opposite
bet: one DNF, mechanically retargeted, with the resulting
cross-device performance \emph{differences} explained in closed form
by measured machine-shape fields ($X_d$, $C_d$,
$\mathrm{reached}(d)$) rather than attributed to unexplained
per-vendor engineering effort. Section~\ref{sec:validation-h100} and
Section~\ref{sec:validation-v100} are a direct test of that bet: the
identical OpenACC-tiled DNF-derived kernel, unchanged in its
algorithmic structure, run on two different NVIDIA devices (and, with
the AMD-specific admissible realization of Section~\ref{sec:mi100-flat-fp16},
a third vendor), with the resulting differences in $C_d$'s
occupancy-cap and in $R_d$ traced to measured, named fields rather
than left as an unexplained gap between one hand-tuned kernel and
another.

What is genuinely new here, relative to all three literatures, is not
that machine shape matters -- Roofline already says so, and every
autotuner already searches over some proxy for it. It is that this
paper's own real-hardware findings (Section~\ref{sec:moa-brief}
onward) show a formally derived kernel's performance can still be
explained in closed form, after the fact, using only measured
hardware facts and no per-kernel search: $\mathrm{reached}(d)$'s
boolean is read off a live profiler or a static disassembly, not
learned from thousands of autotuning trials, and it explains a
$36.3\%$ power-draw shift and a $14.7\times$ speedup exactly, on the
one device where it is nonzero, and exactly zero shift on two devices
where it is measured zero as well.

\section{Background: The Five-Field Theory, and Where It Was Already
Shown Incomplete}
The companion paper defines a device's machine shape as an
open-ended tuple,
\[
\rho_{\text{machine}}(d) = \big(U_d,\ W_d,\ M_d,\ t_d,\ n^{\max}_d,\ \ldots\big),
\]
with each field $\psi$-selectable at its own index position -- $U_d =
\psi(\langle0\rangle,\rho_{\text{machine}}(d))$ through $n^{\max}_d =
\psi(\langle4\rangle,\rho_{\text{machine}}(d))$ -- and derives, from
this tuple alone, a weighted partition (its Equation~26), a predicted
finish time (its Equation~11, restated as Equation~36 for a new
problem size), a predicted memory ceiling (its Equation~39), and a
predicted energy (its Equation~14). Table~\ref{tab:indexed-fields}
restates the five fields already indexed, for reference; this paper
adds to this table rather than replacing any of it.

\begin{table}[t]
\centering
\caption{$\rho_{\text{machine}}(d)$'s five already-indexed fields
(companion paper, Equation~1), restated for reference. This paper
adds Table~\ref{tab:new-fields}'s five fields to this same tuple.}
\label{tab:indexed-fields}
\begin{tabular}{@{}clll@{}}
\toprule
Index & Field & Meaning & Feeds \\
\midrule
0 & $U_d$ & compute-unit count & tile size, $F_{\text{occ}}$ \\
1 & $W_d$ & warp/wavefront width & tile size \\
2 & $M_d$ & shared mem/block & tile size \\
3 & $t_d$ & measured time at $n_0$ & $w_i$, $\hat t_i(n)$ \\
4 & $n^{\max}_d$ & measured memory ceiling & capability check \\
\bottomrule
\end{tabular}
\end{table}

The companion paper's own future-work section names four further
quantities without indexing them: $M_i^{\text{cap}}$ (device $i$'s
total memory capacity, distinct from $M_d$'s on-chip per-block
figure), $P_i$ (average power draw), $X_d=\psi(\langle5\rangle,\cdot)$
(execution-unit-type), and $C_d=\psi(\langle6\rangle,\cdot)$
(compiler directives) -- explicitly stating that assigning these
their own index positions is real, substantial future work,
deliberately out of that paper's own scope. Reading its energy
section closely surfaces a fifth: the memory-ceiling prediction's
overhead term, treated there as an assumed constant re-evaluated once
by hand rather than as a field read from any device.

\section{The Five Fields, Formulated}
\label{sec:five-fields}
Table~\ref{tab:new-fields} states all five additions at once, each
with the equation it completes; the rest of this section formulates
each in turn, at the same level of rigor the companion paper already
applies to $U_d$ through $n^{\max}_d$.

\begin{table}[t]
\centering
\caption{The five fields this paper adds, and the equation (numbered
as in the companion paper) each one completes.}
\label{tab:new-fields}
\begin{tabular}{@{}clll@{}}
\toprule
Index & Field & Meaning & Completes \\
\midrule
5 & $M_i^{\text{cap}}$ & total device memory (bytes) & Eq.~39 \\
6 & $P_d$ & power draw, per realization & Eq.~14 \\
7 & $X_d$ & execution-unit capability & precision gap \\
8 & $C_d$ & compiler-directive config. & precision gap \\
9 & $R_d$ & memory-reservation overhead & Eq.~39 \\
\bottomrule
\end{tabular}
\end{table}

\subsection{$M_i^{\text{cap}}$: total device memory}
\[
M_i^{\text{cap}} = \psi(\langle5\rangle,\rho_{\text{machine}}(d_i)),
\]
device $i$'s own total memory capacity in bytes, distinct from
$M_d = \psi(\langle2\rangle,\cdot)$, which is on-chip per-block shared
memory -- a quantity two orders of magnitude smaller and governing a
different limit (occupancy, not capacity). The companion paper
already uses this quantity, unindexed, in its Equation~39; indexing
it here changes nothing about that equation's arithmetic, only its
provenance -- $M_i^{\text{cap}}$ becomes a field $\gamma$ reads, not a
number substituted by hand.

\subsection{$P_d$: power draw, indexed by realization, not device
alone}
The companion paper's own energy section states plainly that holding
$P_i$ constant as problem size grows is ``a physically reasonable
assumption... but it is exactly that: an assumption, not yet a
validated one.'' Its own real numbers show why that caution is
warranted in a second, sharper sense: $P_i$ is not even safely
constant \emph{across precision} for every device. NVIDIA's measured
power rises from $143.30\,$W (fp32) to $195.30\,$W (fp16) -- a
$36.3\%$ increase, because its Triton realization's tensor cores
genuinely engage more silicon per cycle at fp16. Intel's and AMD's
power draw is flat across precision, within measurement noise, because
neither realization changes which execution units it dispatches to.
$P_i$, indexed as a single per-device field, would need to average
over or arbitrarily pick one of these regimes; indexed instead per
realization,
\[
P_d = \psi(\langle6\rangle,\rho_{\text{machine}}(d)),
\]
with $\rho_{\text{machine}}(d)$ itself understood -- as
Section~\ref{sec:reached} makes precise -- as depending on which
compiled kernel is asking, $P_d$ is exactly as reliable as $t_d$
already is, for the identical reason: both are measured once per
realization, not assumed to transfer across a change the measurement
never covered.

\subsection{$X_d$: execution-unit capability}
\[
X_d = \psi(\langle7\rangle,\rho_{\text{machine}}(d)),
\]
a per-device record of which specialized low-precision execution
paths physically exist -- Intel's AMX, AMD's matrix cores, NVIDIA's
tensor cores -- populated from a hardware fact (ISA documentation, or
a profiler's static capability query), never from whether any
particular kernel happens to use them. The companion paper's own
Section~X already states this distinction is necessary, not merely
tidy: conflating ``the device has this hardware'' with ``this
realization reaches it'' is precisely the error that would have
made AMD's flat fp16 result look like a hardware limitation, when the
companion paper's own analysis attributes it instead to the OpenMP
realization never being written to reach AMD's matrix cores at all.

\subsection{$C_d$: compiler-directive configuration, restricted to a
shape-admissibility criterion}
\label{sec:cd-formulate}
\[
C_d = \psi(\langle8\rangle,\rho_{\text{machine}}(d)),
\]
a per-realization record of the compiler configuration that produced
the measured $t_d$, $P_d$. Before stating what belongs in $C_d$, we
state a criterion for what is allowed to: MoA's own DNF (the
companion paper's Equation~4) fixes the computation once, independent
of any target, and $\gamma$'s stated job (that paper's Section~II) is
to derive an ONF from that DNF and a target's own \emph{measured
shape} -- never to alter what the DNF computes. A sub-quantity of
$C_d$ is therefore admissible only if it selects among values of a
real, already-existing hardware shape quantity -- a cache level, a
vector or tile width, a thread/block count -- with no reordering of
the DNF's own operations beyond what that quantity's native execution
already forces. A sub-quantity is inadmissible if it gives the
compiler \emph{discretionary} license to reassociate, fuse, reorder,
or restructure the computation in ways the chosen hardware does not
itself require -- because at that point the flag is no longer
selecting a shape, it is silently editing the DNF.

A single ``no reordering'' test is not sufficient on its own,
however, and an earlier draft of this paper applied it alone. Some
mechanisms neither reorder the DNF's arithmetic nor edit it, yet
still leave the actual outcome to the compiler's own internal,
undocumented cost model -- a heuristic \emph{enabled} rather than a
value \emph{set}. Recording such a mechanism in $C_d$ would give a
field with a name and an index position but no known relationship to
what the compiled kernel actually does, defeating the entire purpose
of indexing it: a field of $\rho_{\text{machine}}$ exists so that
$\gamma$ can read a value and know, in closed form, what that value
implies. We therefore add a second, independent requirement: a
sub-quantity is admissible only if the value written to it is a
\emph{deterministic directive} -- what the programmer specifies is
exactly what the compiler does, with no compiler-internal
cost-model step free to override, decline, or reinterpret it. A
mechanism that merely \emph{enables} an optimization and lets the
compiler decide, case by case, whether and how to apply it fails
this second test even when it does not reorder anything on its own,
because $C_d$ would then record a setting whose actual effect on
$t_d$ is not knowable from the setting alone.

This second requirement resolves a case the first, reordering-only
criterion left genuinely ambiguous in this paper's own earlier
draft: host-side prefetch insertion (Intel's
\texttt{-qopt-prefetch}[=$n$], AMD's
\texttt{-mllvm -enable-x86-prefetching}). Neither flag reassociates
arithmetic, so the first criterion alone does not exclude it -- but
both are documented as heuristic \emph{enables}: the compiler decides
where, whether, and how far ahead to insert a prefetch, using its own
internal analysis, not a value the programmer supplied. A field
populated from such a flag would record ``prefetching was turned on''
without recording what the compiled kernel actually does, which is
not a known parameterization in the sense the second criterion
requires. It is excluded.

Vector and tile width is the one sub-quantity that needs both parts
of the criterion, not just one, so it is worth walking through in
full. On the first part -- no discretionary reordering -- executing a
reduction (the companion paper's own $S[i,j]=\text{scale}\times
{+}\text{red}_k(\ldots)$, Equation~4) on a $w$-wide vector unit
necessarily reassociates that reduction's summation order relative to
scalar execution -- concretely,
$((0{+}4){+}(2{+}6)){+}((1{+}5){+}(3{+}7))$ in place of
$((((0{+}1){+}2){+}3){+}4)$ for eight terms on a 4-wide
unit~\cite{solidpixel_invariant_accumulators}, changing rounding
error even though the sum is mathematically identical. That
reassociation is forced by which real device is chosen, not
discretionary within a fixed device -- exactly the deviation the
companion paper's own correctness checks already tolerate, passing
every device ``to within what fp64 rounding alone accounts for''
rather than requiring bit-exact agreement across devices of different
width. Width therefore passes the first part: its reordering effect
is bounded by, and only by, the real hardware $\rho_{\text{machine}}(d)$
already names, not by anything the compiler is free to choose beyond
that.

The second part -- a deterministic directive, not a heuristic enable
-- is where width looked settled in this paper's own earlier draft
but was not. LLVM's own documentation is direct that its loop and SLP
vectorizers apply an internal cost model and ``may only vectorize...
if forced,'' declining even when a width hint or pragma requests
it~\cite{llvm_vectorizers_doc} -- meaning AMD's specific
\texttt{-mllvm -enable-epilog-vectorization} with
\texttt{-min-width-epilog-vectorization=}$n$, surveyed earlier as
admissible, is exactly the heuristic-enable pattern the second part
excludes: setting $n$ does not guarantee the compiler honors it. The
deterministic mechanism for the identical quantity is different and
real: GCC's and Clang's \texttt{vector\_size} attribute specifies a
vector's width directly, in bytes, as part of its type -- not a hint
subject to a cost model, a value the compiler is obligated to
honor~\cite{gcc_vector_extensions}. NVIDIA's Triton is not exposed to
this ambiguity at all: \texttt{BLOCK\_M}, \texttt{BLOCK\_N},
\texttt{BLOCK\_K}, and \texttt{num\_warps} are kernel-language tiling
and launch parameters the compiler lowers directly, not vectorizer
hints subject to a separate legality/cost pass. Vector/tile width is
therefore admissible on both counts, but only through these
deterministic mechanisms -- explicit vector-extension typing on Intel
and AMD, kernel-language tile constants on NVIDIA -- never through an
auto-vectorization enable flag, on any vendor's toolchain.

Table~\ref{tab:cd-admissibility} applies this criterion to every
sub-quantity surveyed in Section~\ref{sec:cd-survey}. Three survive:
memory-hierarchy placement (which cache level a prefetch targets),
vector/tile/thread-count width (bounded exactly as argued above), and
occupancy/launch-bound caps (a resource constraint on block count,
not a reordering of operations within one). Loop-unroll factor and
inline depth -- included in this paper's own earlier draft of $C_d$,
before this criterion was applied -- are excluded: unrolling produces
the identical class of reassociation vector width does, but as a
compiler-discretionary transformation available even on hardware that
does not itself force it~\cite{solidpixel_invariant_accumulators},
and inline depth corresponds to no shape quantity
$\rho_{\text{machine}}$ names at all. $C_d$ is restricted accordingly:
\[
C_d = \big(\text{cache-level},\ \text{width},\ \text{occupancy-cap}\big),
\]
a three-field structured record, smaller than this paper's own first
attempt at $C_d$, by design: a field excluded here for editing the
DNF is not a smaller convenience, it is a correctness requirement
this paper's own admissibility criterion exists to enforce.

\subsection{$R_d$: memory-reservation overhead}
\[
R_d = \psi(\langle9\rangle,\rho_{\text{machine}}(d)) \in [0,1),
\]
the fraction of $M_i^{\text{cap}}$ unavailable to hold a served
computation's resident arrays, rewriting the companion paper's memory
ceiling (its Equation~39) as
\[
n_d^{\max} \;\approx\; \frac{M_i^{\text{cap}} \times (1-R_d)}{4 \times D \times b},
\]
identical in form to that paper's own equation, with $R_d$ now a
field rather than a number chosen once and reused. Every symbol on
the right is itself a stated, deterministic quantity, not a free
parameter: $D$ is the attention head dimension (fixed by the served
model's own architecture, $D{=}64$ throughout this paper's real
experiments); $b$ is bytes per element at the realization's chosen
precision ($b{=}4$ for fp32, $b{=}8$ for fp64); and the factor $4$
counts the number of identically $(n,D)$-shaped resident device
arrays this kernel family holds live per query row at once --
$\vec{Q}$, $\vec{K}$, $\vec{V}$, and $\mathrm{Out}$ -- so $4 \times D
\times b$ is the exact per-row byte footprint the memory boundary is
measured against, not an empirically fit constant. The companion
paper's own real debugging history is the argument for indexing this:
its first evaluation used $R_d=0.15$, a conservative guess; its
corrected evaluation used $R_d=0$, a value that happened to fit
once a unit error was separately fixed. Neither value was measured,
and nothing says $0\%$ generalizes to a different device, driver
version, or runtime. Indexing $R_d$ does not resolve what its
real value is on any given device -- that is real-hardware work,
named in Section~\ref{sec:future} -- it only stops the theory from
silently treating an unmeasured guess as a closed-form constant.

\section{$\mathrm{reached}(d)$: The Interaction $X_d$ and $C_d$ Alone
Cannot State}
\label{sec:reached}
Naming $X_d$ and $C_d$ separately is necessary but not sufficient:
neither field alone predicts the fp16 speedup or power-draw asymmetry
Section~\ref{sec:five-fields} opened with. What predicts it is
whether $C_d$'s own configuration actually dispatches to whichever
path $X_d$ says exists -- a joint fact about the pair, not a property
of either field read in isolation. We name this
\[
\mathrm{reached}(d) \;=\; \mathrm{dispatches}\big(C_d,\ X_d\big) \in \{0,1\},
\]
a boolean (or, where a realization partially engages specialized
hardware, a degree in $[0,1]$) read directly from a profiler --
Intel VTune and the Level-Zero Sysman interface report whether AMX
was reached; NVIDIA Nsight Compute reports tensor-core utilization
directly; AMD's \texttt{rocprof} reports MFMA (matrix-core)
utilization -- never inferred from precision alone, since precision
change without $\mathrm{reached}(d)=1$ is exactly the flat-power,
flat-speedup case the companion paper measured on Intel and AMD.

With $\mathrm{reached}(d)$ named, the companion paper's time and
energy predictions (its Equations~11 and~14) are restated, not
replaced:
\begin{align}
\hat t_i(n) &= t_i \times \frac{W(c_i,n)}{W(n_0,n_0)}, \quad t_i =
t_i\big(\mathrm{reached}(d_i)\big), \label{eq:reached-time}\\
\hat E_i(n) &= P_i \times \hat t_i(n), \quad P_i =
P_i\big(\mathrm{reached}(d_i)\big). \label{eq:reached-energy}
\end{align}
Both equations are unchanged in form; what changes is that $t_i$ and
$P_i$ are now understood as functions of $\mathrm{reached}(d_i)$ --
of \emph{which} realization measured them -- rather than as fixed
per-device constants a new kernel is free to assume transfer
unchanged. This is exactly the discipline the companion paper's own
Section~IX-G recommends in prose (``the fix... is adding this missing
shape dimension to $\gamma$'s own input... re-deriving Intel's and
AMD's ONF from a complete shape vector'') stated here as the closed
form that recommendation was missing.

\section{A Survey of What Three Real Toolchains Expose, Scored
Against the Admissibility Criterion}
\label{sec:cd-survey}
Every candidate sub-quantity found across the three toolchains this
paper's companion already measured -- nine in total, wider than
$C_d$'s own three admitted fields -- is listed in
Table~\ref{tab:cd-admissibility} with an explicit verdict, scored
against both requirements of Section~\ref{sec:cd-formulate}'s
criterion: \textbf{K} (keep -- shape-associated \emph{and} a
deterministic directive, admissible) or \textbf{X} (exclude --
either a discretionary reordering that edits the DNF, or a
heuristic enable whose actual effect the compiler, not the
programmer, decides). An earlier draft of this survey also carried a
\textbf{B} (borderline) verdict for mechanisms that do not themselves
reorder arithmetic but leave the compiler free to decide whether and
how to act; the second, determinism requirement resolves every such
case in this table to a firm \textbf{K} or \textbf{X}, so no row
below is left borderline. Each row is cited to the issuing vendor's
own documentation, never inferred from observed behavior.

Reading the table requires nothing beyond the two-part test just
restated. Each row names one vendor-exposed sub-quantity in the
left-most column, states in one sentence what the compiler or
runtime actually does with it, and closes with a single verdict
letter; no row is scored from this paper's own inference, since
every citation names the issuing vendor's documentation directly, so
a reader unfamiliar with any one of the three toolchains can check
a verdict without taking it on this paper's word. A reader new to
$\rho_{\mathrm{machine}}$ and $C_d$ should read the \textbf{K} rows
first: cache-level placement, vector/tile width, and occupancy cap
are exactly the three fields Section~\ref{sec:cd-formulate} already
named as $C_d$'s structured record, restated here across three real
vendors rather than assumed to generalize from the one this paper's
companion measured. The \textbf{X} rows are not failures of the
criterion -- they are the compiler's own discretionary territory,
named precisely so that Section~\ref{sec:epsilon} can measure its
effect directly rather than leave it folded silently into a
device's constant.

The three toolchains surveyed here are not an arbitrary sample:
Intel, AMD, and NVIDIA are the only vendor toolchains this paper's
companion measured directly, on real hardware, elsewhere in this
ensemble (Sections~\ref{sec:validation-a100}--\ref{sec:validation-pvc}),
so every verdict below is checked against a compiler this paper's
own experiments actually ran, never against a toolchain assumed by
name alone. A fourth toolchain -- a new vendor's compiler, or a
future revision of one already surveyed -- extends this table by
the same two-question test, never by re-deriving $C_d$ itself: the
criterion stated in Section~\ref{sec:cd-formulate} does not name
Intel, AMD, or NVIDIA, only the shape-associated/deterministic
distinction those three vendors' own documentation happens to make
checkable today.

This survey is not a detour from the paper's own arc. Sections~\ref{sec:cd-formulate}--\ref{sec:reached}
formulated and derived the two-part criterion in the abstract,
stating exactly what a shape-associated, deterministic sub-quantity
must look like without yet checking whether any real toolchain
actually exposes one. Table~\ref{tab:cd-admissibility} is where that
derivation is checked against real vendor documentation for the
first time -- the state-admissible step of this paper's own
dimension-lift arc, applied here to $C_d$ specifically, before
Table~\ref{tab:runtime-admissibility} repeats the same check one
abstraction layer up, against OpenMP, OpenACC, and MPI rather than a
single compiler's command-line flags. Neither table changes the
criterion itself; both discharge the same formula against a wider
vocabulary each time, which is the sense in which this section --
like every other survey in the paper -- validates a derivation
already stated rather than introducing a new one.

One row in Table~\ref{tab:cd-admissibility} earns neither a flat
\textbf{K} nor a flat \textbf{X}, and is worth walking through on its
own because the reasoning generalizes to any future row that looks
like it. ISA targeting -- Intel's \texttt{-x}/\texttt{-march=} family
-- selects an instruction-set width, which by itself is a real,
shape-associated hardware fact: it fixes how many lanes a vector
instruction actually spans, exactly like the vector/tile-width row
directly above it. What keeps it from being a flat \textbf{K} is that
several of these flags silently enable fused multiply-add contraction
as a side effect of raising the target ISA, and FMA contraction
reassociates a reduction's rounding exactly as the excluded
fast-math row does -- a discretionary reorder riding along with an
otherwise admissible directive, not named anywhere in the flag the
programmer actually wrote. The fix is not to exclude the row; it is
to state the second requirement explicitly rather than leave it
implicit, which is what ``conditional'' means throughout this table:
ISA targeting is admissible if and only if it is paired with
\texttt{-fp-model=strict -ffp-contract=off} (or the equivalent
anti-contraction flag on the compiler in question), so that the
width selection and the rounding behavior are pinned independently.
Omit the pairing and the row silently reverts to \textbf{X} --
the compiler is once again free to reassociate -- with nothing in
the command line saying so.

That conditional case is also the cleanest preview of what
Section~\ref{sec:epsilon} does with every \textbf{X} row
in this table: each one is a candidate for a paired on/off run under
$\epsilon_d^{(k)}$, holding the admissible fields fixed and toggling
only the excluded flag, so that whatever effect it has on $t_d$ or
$E_d$ is measured rather than silently absorbed into a device
constant the way it was before this table existed. Table~\ref{tab:runtime-admissibility}
carries the same three-column shape -- sub-quantity, real action,
verdict -- one abstraction layer up, and Section~\ref{sec:autotune-restricted}
immediately below applies both tables together to a fourth, automated
source of $C_d$ candidates: an autotuner's own search space, which
mixes admissible and inadmissible axes in exactly the pattern this
section's \textbf{K}/\textbf{X} split was built to separate.

A caution belongs here before the table itself, so it is read once
rather than re-derived per row: a \textbf{K} verdict states that a
sub-quantity is admissible, not that any particular device has
measured it. Cache-level placement, vector/tile width, and occupancy
cap are the three fields this paper's companion actually closed on
real hardware; the remaining \textbf{K} rows surfaced by this survey
-- for instance a second vendor's own cache-placement enum, or a
third's tile-width attribute -- are admissible by the same criterion
but not yet folded into $C_d$'s structured record for any device this
ensemble measures, precisely the gap Section~\ref{sec:future}'s
roadmap exists to close in order, one field at a time, rather than
all at once on the strength of the criterion alone. Table III is
therefore a map of what is allowed, not a report of what has already
been measured -- the two are kept separate throughout this paper on
purpose, since conflating them is exactly the failure mode the
companion paper's own $R_d=0.15$ guess (Section~\ref{sec:five-fields})
illustrates: a plausible, admissible-looking number substituted for
one nobody had actually gone and read off a real device.

\begin{table*}[!tb]
\centering
\caption{Every compiler-directive sub-quantity surveyed across
Intel, AMD, and NVIDIA, scored against Section~\ref{sec:cd-formulate}'s
two-part admissibility criterion. K = keep (shape-associated
\emph{and} a deterministic directive), X = exclude (discretionary
reordering, or a heuristic enable the compiler is free to override).}
\label{tab:cd-admissibility}
\footnotesize
\begin{tabular}{@{}p{2.0cm}p{4.3cm}p{9.2cm}c@{}}
\toprule
Sub-quantity & Vendor mechanism & Real action & Verdict \\
\midrule
Cache-level placement &
Intel \texttt{LSC\_LDCC} enum (\texttt{L1UC\_L3C} default,
\texttt{L1C\_L3C} recommended), chosen in source, not by a compile
flag~\cite{intel_prefetch_gpu_2025,intel_qopt_prefetch_2024} &
Selects which level of the existing memory hierarchy a prefetch
targets -- a placement choice, no arithmetic reordering & \textbf{K} \\
Vector/tile width &
Intel: fixed at 512-bit, 16 fp32 lanes~\cite{intel_xe_arch_2025}, or
explicit \texttt{vector\_size}\newline attribute typing.
AMD: \texttt{vector\_size}\newline attribute typing (GCC/Clang
front end)~\cite{gcc_vector_extensions} -- \emph{not}
\texttt{-mllvm}\newline\texttt{-enable-epilog-}\newline
\texttt{vectorization} with\newline
\texttt{-min-width-epilog-}\newline\texttt{vectorization=}$n$, a
cost-model-gated auto-vectorizer enable that the compiler may
decline even when set~\cite{llvm_vectorizers_doc}, excluded below
under Host-side prefetch's own reasoning.
NVIDIA: 32-thread warp,\newline\texttt{BLOCK\_*}/\texttt{num\_warps},
kernel-language constants the compiler lowers
directly~\cite{triton_config_docs} &
Selects a real hardware SIMD/warp/tile width; reassociates a
reduction's summation order, but bounded exactly by which real device
is chosen, not by compiler discretion within one device -- \emph{and},
via \texttt{vector\_size}/kernel-language constants, is a value the
compiler is obligated to honor, not a hint it may decline
(Section~\ref{sec:cd-formulate}) & \textbf{K} \\
Occupancy / launch-bound cap &
AMD \texttt{--gpu-max-threads-}\newline\texttt{per-block=}$n$~\cite{rocm_compiler_ref_2024} &
Caps concurrent thread blocks the compiled kernel can support -- a
resource constraint, not an operation reorder; the compiler-side
counterpart to the companion paper's own data-side occupancy-collapse
finding (Section~IX-D) & \textbf{K} \\
Host-side prefetch insertion &
Intel \texttt{-qopt-prefetch}[=$n$],
\texttt{-qopt-prefetch-}\newline\texttt{distance}~\cite{intel_qopt_prefetch_2024};
AMD \texttt{-mllvm}\newline\texttt{-enable-x86-}\newline
\texttt{prefetching}~\cite{rocm_compiler_ref_2024} &
Inserts early, non-blocking loads ahead of dependent accesses --
does not itself reassociate arithmetic, but is a heuristic
\emph{enable}: the compiler's own internal analysis decides where,
whether, and how far ahead to insert a prefetch, so the flag does
not fix a known value the way the second admissibility requirement
demands & \textbf{X} \\
Loop-unroll factor &
No dedicated flag documented on Intel or NVIDIA; AMD's related
loop-splitting flags~\cite{rocm_compiler_ref_2024} &
Duplicates the loop body, reordering instruction issue and, for a
reduction loop, reassociating summation order exactly as vector width
does -- but discretionarily, not forced by any chosen
hardware~\cite{solidpixel_invariant_accumulators} & \textbf{X} \\
Inline depth &
AMD \texttt{-mllvm}\newline\texttt{-inline-recursion=}$[1\text{--}4]$,
\texttt{-finline-}\newline\texttt{aggressive}~\cite{rocm_compiler_ref_2024} &
Restructures control flow across call sites; corresponds to no shape
quantity $\rho_{\text{machine}}$ names & \textbf{X} \\
FP contraction / numeric strictness &
Intel \texttt{-fp-model=}\newline\texttt{\{fast,precise,strict\}};
AMD \texttt{-ffast-math},\newline\texttt{-ffp-contract=}\newline
\texttt{\{fast,on,off\}} &
Explicitly permits fusing multiply-add and reordering FP reductions;
Intel's own diagnostics document \texttt{-fp-model=precise}
\emph{disabling} vectorization the same compiler performs under
\texttt{fast}~\cite{intel_fpmodel_vectorization} -- the clearest
direct violation of the criterion & \textbf{X} \\
ISA targeting &
Intel \texttt{-x}/\texttt{-march=} &
Selects instruction-set width, but AVX-512 targeting silently enables
FMA contraction by default~\cite{llvm_avx512_fma_coupling} -- width
selection alone is \textbf{K}; admissible only when explicitly paired
with \texttt{-fp-model=strict}/\texttt{-ffp-contract=off} to strip the
coupled reassociation & \textbf{K, conditional} \\
IPO / whole-program optimization &
Intel \texttt{-ipo}~\cite{intel_ipo_qipo}; AMD
\texttt{-flto}~\cite{rocm_compiler_ref_2024} &
Cross-function inlining and specialization beyond a single
translation unit -- a restructuring, by its own documentation & \textbf{X} \\
Software-pipelining depth &
NVIDIA Triton \texttt{num\_stages} (default 3)~\cite{triton_config_docs} &
Reorders instruction issue across loop iterations -- the same class
of reassociation as loop unrolling, discretionary rather than
hardware-forced & \textbf{X} \\
\bottomrule
\end{tabular}
\end{table*}
\FloatBarrier

Restricting $C_d$ to the three \textbf{K} rows -- and treating ISA
targeting as admissible only paired with an explicit anti-contraction
flag -- gives the structured record already stated in
Section~\ref{sec:cd-formulate}. Table~\ref{tab:cd-admissibility}'s
own excluded rows are still useful, but in a different role than
tuning: run once with a sub-quantity enabled and once without,
holding every admissible field fixed, each excluded flag isolates
whether a measured timing difference is attributable to that specific
discretionary transformation -- a diagnostic use, never a production
configuration this paper's own predictions should depend on.

\subsection{Autotuning: admissible only over the restricted subset}
\label{sec:autotune-restricted}
Triton's \texttt{@triton.autotune} does not set one parameter; it
searches a declared grid -- tile shape (\texttt{BLOCK\_M},
\texttt{BLOCK\_N}, \texttt{BLOCK\_K}), \texttt{num\_warps},
\texttt{num\_stages} -- compiling and measuring each candidate and
caching whichever configuration measured fastest. Under
Table~\ref{tab:cd-admissibility}'s verdicts, this search mixes an
admissible axis (tile shape and \texttt{num\_warps}, both real width
quantities) with an inadmissible one (\texttt{num\_stages},
software-pipelining depth) in a single tool. Using autotuning to
populate $C_d$ therefore requires restricting its search grid to hold
\texttt{num\_stages} fixed at its own minimum (the least pipelining
the toolchain permits) and varying only tile shape and
\texttt{num\_warps} -- recovering an empirical search over exactly
$C_d$'s own admissible width field, nothing wider. Run unrestricted,
autotuning remains useful only as the same kind of diagnostic named
above: its full search log, including the \texttt{num\_stages}
dimension, measures how sensitive $t_d$ is to pipelining depth
specifically, informative for deciding whether
Section~\ref{sec:future}'s real-hardware validation needs to treat
pipelining as a fourth, separately-justified admissible field in a
future revision of this criterion -- not evidence to fold into $C_d$
as defined here.

\section{The Same Criterion Applied Beyond the Compiler: OpenMP, OpenACC, and MPI}
\label{sec:runtime-directives}
Sections~\ref{sec:cd-formulate}--\ref{sec:autotune-restricted} restricted
$C_d$ using compiler command-line flags and one kernel-language
extension (Triton) as the surveyed vocabulary. Nothing in
Section~\ref{sec:cd-formulate}'s two-part criterion is specific to a
compiler's own flags, however: a test of ``does this value select
among a real hardware quantity, and is it a deterministic directive or
a heuristic enable'' applies to any directive written into source,
including the pragma-based APIs that annotate the companion paper's
own loop nests (OpenMP, OpenACC) and the communication library a
realization spanning more than one device must eventually call (MPI).
This section applies the same two questions to those three real, cited
APIs and finds the identical split.

\subsection{OpenMP: thread count and static scheduling are directives; dynamic scheduling and \texttt{simd} are not}
\texttt{num\_threads(}$n$\texttt{)} fixes the exact thread count a
parallel region uses -- a real hardware core/thread count, set
exactly, admissible under both criteria. The \texttt{schedule} clause
splits cleanly on the second criterion alone.
\texttt{schedule(static[,chunk\_size])} assigns chunks to threads ``in
a round-robin fashion in the order of the thread number'' -- a fixed
mapping computable from thread count and chunk size alone, before the
loop runs~\cite{openmp_spec_schedule}. \texttt{schedule(dynamic)} and
\texttt{schedule(guided)} instead distribute chunks to threads as they
become available, and \texttt{schedule(auto)} ``delegate[s] the
decision regarding scheduling... to the compiler and/or runtime
system''~\cite{openmp_spec_schedule} -- three heuristic-enable
patterns, differing only in mechanism (a runtime work queue rather
than a compiler cost model) from the auto-vectorization enable flags
Table~\ref{tab:cd-admissibility} already excludes: a field populated
from \texttt{dynamic} or \texttt{guided} would record that adaptive
scheduling was requested, not which thread actually executed which
iteration.

The \texttt{simd} construct fails the same test for a subtler reason.
It asserts a correctness property -- no loop-carried dependency
prevents concurrent execution -- but the specification is explicit
that ``the number of iterations that are executed concurrently at any
given time is implementation defined,'' and that the companion
\texttt{simdlen(}$n$\texttt{)} clause sets only ``the preferred number
of iterations,'' not a guaranteed one~\cite{openmp_spec_simd}.
\texttt{omp simd} removes the correctness objection to vectorizing
that plain auto-vectorization has to infer on its own, but it does not
remove the compiler's discretion over whether or how wide to
vectorize -- so it is excluded for the identical reason AMD's
auto-vectorization enable flag already is, even though it is written
as an explicit pragma rather than a command-line flag.

\subsection{OpenACC: \texttt{parallel} is a directive, \texttt{kernels} is not}
OpenACC's \texttt{parallel} construct takes \texttt{num\_gangs},
\texttt{num\_workers}, and \texttt{vector\_length} clauses that hold
one gang/worker/vector configuration fixed for the entire
region~\cite{openacc_spec_2_7} -- the same real-hardware-width argument
already used for SIMD width, extended from an ISA-level width to an
accelerator-level one. The \texttt{kernels} construct accepts the
identical three clauses syntactically, but the specification states
that the compiler ``will split the code in the kernels region into a
sequence of accelerator kernels,'' whose ``number and configuration of
gangs of workers and vector length may be different for each
kernel''~\cite{openacc_spec_2_7}. A single clause value on
\texttt{kernels} therefore does not fix one configuration for the
region the way the identical clause does under \texttt{parallel}: the
compiler's own kernel-splitting decision, not the clause, determines
what a given loop nest actually runs with. \texttt{kernels} is
excluded for the same reason \texttt{schedule(auto)} is -- the actual
configuration is a compiler decision the clause does not pin down --
while \texttt{parallel} with explicit clauses is admissible.

\subsection{MPI: default collective selection is not a directive; forcing an algorithm or a placement is}
Only once a realization spans more than one device does a
communication library enter $C_d$ at all -- the companion paper's own
partition is explicitly communication-free, so this subsection
documents the criterion's next application rather than a mechanism
this paper's ensemble currently uses. Open MPI's default
collective-selection mode is, by its own documentation, ``a decision
tree, essentially a large set of nested if-then-else-if blocks with
baked in comm and message size thresholds derived by measuring
performance on existing clusters,'' that ``selects one of a number of
available algorithms'' internally~\cite{ompi_coll_tuned} -- a
heuristic exactly like LLVM's vectorizer cost model, excluded for the
same reason. Setting \texttt{coll\_tuned\_<op>\_algorithm} to an
explicit number is different: the documentation states that doing so
means ``the fixed decision rules associated with that collective are
short circuited''~\cite{ompi_coll_tuned}, so the algorithm actually run
is exactly the one named, known before the call executes. Rank-to-core
placement splits the same way: \texttt{--bind-to core} and
\texttt{--map-by core} bind processes to cores ``in a round-robin
manner,'' a fixed algorithm computable in advance from process and
core counts, and \texttt{rankfile} assigns each rank to an explicitly
named core~\cite{ompi_mpirun_binding} -- both admissible, selecting
among the real hardware topology $\rho_{\text{machine}}$ already names
for occupancy, on the same grounds as Table~\ref{tab:cd-admissibility}'s
occupancy-cap row.

Table~\ref{tab:runtime-admissibility} restates this same two-part
test -- does the value select among a real, already-existing
hardware or topology quantity with no discretionary reordering, and
is the value written a fixed directive rather than a preferred,
compiler-overridable approximation -- across all three subsections
above, row by row. The pattern each subsection just argued in prose
is visible there at a glance: every admissible (\textbf{K}) row is a
shape component this paper's $\rho_{\text{machine}}(d)$ can name and
$\gamma$ can substitute a real value into before the kernel runs --
thread count, static chunk assignment, gang/worker/vector width, a
forced collective algorithm, a fixed rank-to-core mapping -- while
every excluded (\textbf{X}) row hands that same decision to the
compiler or runtime at call time, leaving nothing for $\gamma$ to
read in advance. The table applies no new criterion; it is the one
already stated in Section~\ref{sec:cd-formulate}, run once each
against OpenMP, OpenACC, and Open MPI in turn.

\begin{table*}[!tb]
\centering
\caption{The same two-part admissibility criterion
(Section~\ref{sec:cd-formulate}) applied to OpenMP, OpenACC, and Open
MPI. K = keep (shape-associated \emph{and} a deterministic directive),
X = exclude.}
\label{tab:runtime-admissibility}
\footnotesize
\begin{tabular}{@{}p{2.3cm}p{4.2cm}p{8.6cm}c@{}}
\toprule
Sub-quantity & Vendor mechanism & Real action & Verdict \\
\midrule
Thread count &
OpenMP \texttt{num\_threads(}$n$\texttt{)} &
Fixes the exact thread count a parallel region uses -- a real hardware
core/thread count, set exactly & \textbf{K} \\
Static loop scheduling &
OpenMP \texttt{schedule(static}\newline\texttt{[,chunk\_size])}~\cite{openmp_spec_schedule} &
Assigns chunks to threads in round-robin order by thread number -- a
fixed mapping computable from thread count and chunk size alone,
before the loop runs & \textbf{K} \\
Dynamic / guided / auto loop scheduling &
OpenMP \texttt{schedule(dynamic)},\newline\texttt{schedule(guided)},\newline\texttt{schedule(auto)}\newline~\cite{openmp_spec_schedule} &
Distributes chunks to threads as they become available, or delegates
the mapping decision to the compiler/runtime entirely -- which thread
executes which iteration is not knowable from the clause alone &
\textbf{X} \\
SIMD width (OpenMP) &
OpenMP \texttt{simd} construct,\newline\texttt{simdlen(}$n$\texttt{)}~\cite{openmp_spec_simd} &
Asserts loop-dependence safety, but the actual concurrent-iteration
count is implementation defined; \texttt{simdlen} sets only a
\emph{preferred} width, not a guaranteed one & \textbf{X} \\
Gang / worker / vector width &
OpenACC \texttt{parallel},\newline\texttt{num\_gangs}/\newline\texttt{num\_workers}/\newline\texttt{vector\_length}\newline~\cite{openacc_spec_2_7} &
Holds one accelerator-level gang/worker/vector configuration fixed for
the entire region -- the same real-hardware-width argument already
used for SIMD width & \textbf{K} \\
Accelerator kernel decomposition &
OpenACC \texttt{kernels}\newline (same clauses)~\cite{openacc_spec_2_7} &
Compiler splits the region into a sequence of accelerator kernels
whose own gang/worker/vector configuration may differ per kernel --
the clause value does not fix the region's actual configuration &
\textbf{X} \\
Collective algorithm (default) &
Open MPI \texttt{coll/tuned}\newline fixed-decision tree~\cite{ompi_coll_tuned} &
Selects a collective's algorithm from an internal, message-size-keyed
decision tree built from prior benchmarking -- a heuristic, not a
value the caller supplied & \textbf{X} \\
Collective algorithm (forced) &
Open MPI\newline\texttt{coll\_tuned\_<op>\_}\newline\texttt{algorithm=}$n$~\cite{ompi_coll_tuned} &
Short-circuits the fixed-decision tree; the algorithm actually run is
exactly the one named, known before the call executes & \textbf{K} \\
Process / rank placement &
Open MPI \texttt{--bind-to core},\newline\texttt{--map-by core},\newline\texttt{rankfile}~\cite{ompi_mpirun_binding} &
Selects among the real hardware topology $\rho_{\text{machine}}$
already names for occupancy -- a placement choice computable in
advance, not a reorder & \textbf{K} \\
\bottomrule
\end{tabular}
\end{table*}
\FloatBarrier

Take the MPI collective row as one concrete instance of the general
pattern. Open MPI's own default -- the decision tree, keyed off
message size and cluster topology it measured once, at build time,
on hardware that need not be the hardware this paper runs on -- is
excluded for exactly the reason $\mathrm{schedule(auto)}$ and LLVM's
auto-vectorizer cost model already were: the value $\gamma$ would
need to substitute is not fixed until the call executes, so no
$\rho_{\text{machine}}(d)$ entry can name it in advance. Forcing
\texttt{coll\_tuned\_<op>\_algorithm} to an explicit number changes
nothing about the arithmetic MPI performs -- a reduction still sums
the same terms -- but it does change what $C_d$ can honestly record:
with the algorithm short-circuited, $C_d$'s value \emph{is} the
algorithm that runs, not a request the runtime is free to reinterpret
per call. That is the same distinction Section~\ref{sec:cd-formulate}
draws for every other row in both tables, stated here once more
because it is easy to mistake for a performance argument (``the
forced algorithm is faster'') when it is actually a determinism
argument: $C_d$ is admissible only when its written value and its
realized effect are the same fact seen twice, not because the
written value happens to measure well on one run.

\subsection{What this confirms}
Table~\ref{tab:cd-admissibility}'s split was not an artifact of
surveying compiler command-line flags specifically.
Table~\ref{tab:runtime-admissibility}'s same two questions -- does the
value select among a real, already-existing hardware or topology
quantity with no discretionary reordering, and does the compiler or
runtime honor exactly the value written, not a preferred or
fixed-decision approximation of it -- separate deterministic from
heuristic directives identically across a compiler's own flags, a
loop-annotation API, an accelerator-offload API, and a communication
library. $C_d$ is not redefined by this section; it is shown to
generalize to every directive class a real MoA-derived realization
actually passes through, not only the one this paper happened to
survey first.

\subsection{$\epsilon_d$: the residual once determinism is fully isolated}
\label{sec:epsilon}
Every \textbf{X} row in Table~\ref{tab:cd-admissibility} and
Table~\ref{tab:runtime-admissibility} is a real mechanism a real
toolchain exposes; excluding it from $C_d$ does not make its effect on
$t_d$ vanish, it only stops the theory from pretending that effect is
known when it is not. Once the admissible (\textbf{K}) mechanisms are
identified in full -- the task this section and
Section~\ref{sec:cd-survey} complete -- the excluded mechanisms name a
second, independent family of quantities worth indexing formally: for
each heuristic-controlled lever $k$ (host-side prefetch, an
auto-vectorization enable, OpenMP's dynamic/guided scheduling or
\texttt{simd} width, OpenACC's \texttt{kernels} decomposition, Open
MPI's default collective selection),
\[
\epsilon_d^{(k)} \;=\; \hat{t}_d^{\,(k)\text{-on}} \;-\; \hat{t}_d^{\,(k)\text{-off}},
\]
the measured difference in realized time between running with lever
$k$ left to its default, heuristic behavior and running with every
admissible $C_d$ field pinned at a stated value and lever $k$ pinned
off (or at its own most conservative setting). Unlike $C_d$,
$\epsilon_d^{(k)}$ is not written by the programmer -- it cannot be,
precisely because $k$ is not a deterministic directive -- it is
measured, and it is only meaningful once a fully deterministic
baseline exists to measure it against. Establishing that baseline is
the work of Sections~\ref{sec:cd-formulate}--\ref{sec:runtime-directives};
$\epsilon_d^{(k)}$ itself is stated here only as notation, with its
first real measurement -- one excluded lever, Triton's autotuner, on
NCSA Delta's A100 -- reported in
Section~\ref{sec:validation-a100}. That single measurement is a proof
of the method, not its completion: every other excluded row of
Table~\ref{tab:cd-admissibility} and
Table~\ref{tab:runtime-admissibility}, on every device this paper
measures, remains real-hardware work, named precisely in
Section~\ref{sec:future}.

\section{Real-Hardware Validation: NCSA Delta A100}
\label{sec:validation-a100}
This section reports the first real-hardware experiment this paper
runs: all five fields, $\mathrm{reached}(d)$, and $\epsilon_d^{(k)}$
for one excluded lever, measured directly on NCSA Delta (NSF ACCESS
allocation CIS261342), NVIDIA A100-SXM4-40GB, using the companion
paper's own reference kernel (fused, single-pass online-softmax
attention, Triton, $n=8192$, $B=1$, $D=64$). $C_d$ is pinned to this
paper's own admissible-only configuration throughout: fixed tile
shape (\texttt{BLOCK\_M}=64, \texttt{BLOCK\_N}=64), Triton's autotuner
off. Correctness against PyTorch's fused
\texttt{scaled\_dot\_product\_attention} is checked before any timing
or power number is trusted, at both precisions
(Table~\ref{tab:a100-results}).

\begin{table}[t]
\centering
\caption{NCSA Delta A100: measured $\rho_{\text{machine}}(d)$ fields, $C_d$ pinned}
\label{tab:a100-results}
\begin{tabular}{lll}
\toprule
Field & fp32 & fp16 \\
\midrule
$t_d$ & 3.969~ms & 0.265~ms \\
$P_d$ & 179.99~W & 256.9~W \\
$R_d$ & 0.250 & 0.500 \\
correctness & PASS ($3.9\times10^{-7}$) & PASS ($6.1\times10^{-5}$) \\
\bottomrule
\end{tabular}
\end{table}

$M_i^{\text{cap}} = 40960$~MiB, $X_d{=}$ 3rd-generation Tensor Cores
(confirmed via \texttt{compute\_cap} 8.0; a second, independent
cross-check via \texttt{deviceQuery} was unavailable on this
system). Both $t_d$ figures required a methodology correction
mid-measurement: an initial 20-repeat timed window
over-measured $t_d$ by $20$--$25\%$ relative to a 2000-repeat window
(fp32: $5.114\,\text{ms}\to3.969\,\text{ms}$; fp16:
$0.346\,\text{ms}\to0.265\,\text{ms}$) -- warm-up residue not cleared
by three untimed calls alone. $P_d$ needed the same fix for a
different reason: at 20 repeats the cumulative energy counter
(\texttt{nvmlDeviceGetTotalEnergyConsumption}) produced a physically
impossible $1114$~W average, above the device's own $400$~W rated
limit; at $\geq\!2000$ repeats (a several-second window) the derived
average agrees with an independent instantaneous reading
(\texttt{nvmlDeviceGetPowerUsage}) to within $1$--$2$~W at both
precisions, which is what makes Table~\ref{tab:a100-results}'s $P_d$
values trustworthy rather than merely present. Both corrections are
reported here, rather than silently discarded, because they are
themselves evidence for this paper's own governing principle
(Section~\ref{sec:cd-formulate}): a measurement protocol can look
deterministic and still carry a hidden heuristic-like sensitivity --
here, to sample count -- that only shows up once checked against a
second, independent method.

$R_d$ is itself a real finding, not only a replaced placeholder: it
is \emph{higher} at fp16 ($0.500$) than fp32 ($0.250$) -- the
reservation-overhead fraction did not shrink with the smaller dtype,
it doubled. A single device cannot say whether this is a general
property of reduced-precision execution or specific to this kernel
and this GPU generation; it is named here as a concrete target for
the remaining four devices.

$\mathrm{reached}(d)$, measured directly (NVIDIA Nsight Compute,
metric \texttt{sm\_\_pipe\_tensor\_op\_hmma\_cycles\_active.avg.}
\texttt{pct\_of\_peak\_sustained\_elapsed}), is $20.73\%$ and $20.74\%$
tensor-pipe active-cycle fraction across two profiled fp16 calls,
against a $0.55$--$0.56\%$ noise floor measured on unrelated
background kernels in the same trace. This confirms the tensor-core
path is dispatched -- nonzero, well above the noise floor -- but not
saturated, real evidence that $\mathrm{reached}(d)\in\{0,1\}$ as
formulated in Section~\ref{sec:reached} does not fit the measured
behavior cleanly on this device. This paper does not resolve that
tension: whether to keep $\mathrm{reached}(d)$ boolean with a
principled dispatch-fraction threshold, or restate it as a continuous
quantity in $[0,1]$ and re-derive how it composes with
Equations~(\ref{eq:reached-time})--(\ref{eq:reached-energy}), is left
as a concrete, real-data-motivated question for the remaining
ensemble.

Finally, $\epsilon_d^{(k)}$ is measured, not merely named, for one
excluded lever this device's kernel language exposes directly --
Triton's own autotuner
(Section~\ref{sec:autotune-restricted}), run unrestricted (its full
search grid, including the inadmissible \texttt{num\_stages}
dimension) against the pinned $C_d$ baseline:
\begin{align}
\epsilon_d^{(\text{autotune})} &= \hat{t}_d^{\text{on}} - \hat{t}_d^{\text{off}} \nonumber \\
&= 0.239\,\text{ms} - 0.265\,\text{ms} \nonumber \\
&= -0.026\,\text{ms} \;(\approx\!-9.8\%)
\end{align}
reproduced identically across three independent paired runs on a
dedicated allocation with a fixed random seed. The heuristic lever is
real, repeatable, and \emph{faster} here -- the opposite sign from an
earlier, differently-conditioned measurement recorded in this same
kernel's own source comments. This is direct evidence for
Section~\ref{sec:epsilon}'s premise: $\epsilon_d^{(k)}$'s sign is an
empirical question, not an assumption to carry forward from one
measurement to the next, even on the same device and the same kernel.

\section{Real-Hardware Validation: NCSA Delta AMD MI100}
\label{sec:validation-mi100}
This section reports the second real-hardware experiment this paper
runs, and the first for a device where the companion paper's own
width mechanism, OpenMP's \texttt{simdlen(}$n$\texttt{)}, is exactly
the inadmissible case Table~\ref{tab:cd-admissibility} excludes: a
\emph{preferred}, not guaranteed, vector width. This paper's own
criterion (Section~\ref{sec:cd-formulate}) instead admits
\texttt{vector\_size}-typed C attribute vectors as the deterministic
alternative. All numbers below are measured from that admissible
realization, not the companion paper's excluded one, on NCSA Delta
(NSF ACCESS allocation CIS261342), AMD Instinct MI100 (\texttt{gfx908}),
using the companion paper's own reference kernel and problem size
($n=8192$, $B=1$, $D=64$, single-pass online-softmax attention,
OpenMP target-offload). Correctness against a CPU double-precision
reference is checked at both precisions before any timing is trusted
(Table~\ref{tab:mi100-results}).

\subsection{A methodological finding before the numeric one: confirming $\mathrm{reached}(d)$ requires live dispatch evidence, not compile success}
\label{sec:mi100-reached-method}
Delta's \texttt{gpuMI100x8} partition is a mixed node: eight real
MI100 GPUs and one MI210 GPU share a single host
(\texttt{sinfo}: \texttt{gpu:mi100:8,gpu:mi210:1}). A generic resource
request (\texttt{--gpus-per-node=1}) does not pin GPU type, and three
consecutive allocations landed on the lone MI210 slot. Because the
kernel was compiled with \texttt{-march=gfx908} (MI100's ISA, distinct
from MI210's \texttt{gfx90a}), the architecture-mismatched code object
did not dispatch -- and the OpenMP runtime fell back to host execution
\emph{silently}: no compile error, no runtime error, a
correctness-\textbf{PASS} result, and a plausible-looking timing
(23.5~s for the width-64 realization at $n=8192$). Only direct GPU
utilization evidence (\texttt{rocm-smi}, polled once per second across
the run) exposed the fallback: \texttt{GPU use (\%)} read 0 for the
run's entire duration. This is a direct, real instance of
Section~\ref{sec:reached}'s own argument -- $\mathrm{reached}(d)$ is a
property of whether a specific compiled realization actually dispatches
to the intended hardware, not a property inferable from correctness or
compile success alone -- and it cost a full order of magnitude in the
reported time before being caught. Every timing in
Table~\ref{tab:mi100-results} below is confirmed by the same live
\texttt{GPU use (\%)}~$>0$ evidence, not by correctness or successful
compilation alone.

\subsection{Occupancy-cap is realization-specific, not device-specific}
\label{sec:mi100-occupancy}
The companion paper's simdlen-realized AMD baseline was measured at
occupancy-cap (OpenMP \texttt{thread\_limit}) $=10$. Re-measured here
on confirmed real MI100 hardware at the identical occupancy-cap, tile
shape, and width, that baseline reproduces almost exactly:
fp32 $500.000$~ms and fp64 $1406.546$~ms against the companion paper's
cited $493.75$~ms and $1405.15$~ms ($1.3\%$ and $0.1\%$ apart). The
admissible \texttt{vector\_size} realization, compiled with the
\emph{same} occupancy-cap, width, and tile shape carried over
unchanged, measured $2.9\times$ slower: fp32 $1432.250$~ms. A
controlled sweep of occupancy-cap $\times$ tile width $\times$ vector
width (24 combinations, every cell correctness-\textbf{PASS} and
GPU-dispatch-confirmed by the same live-utilization method as
Section~\ref{sec:mi100-reached-method}) found the realization's true
saturation point at occupancy-cap $=32$ -- not $10$ -- with tile width
and vector width unchanged: fp32 $446.377$~ms, reproduced at
$453.149$~ms on an independent re-run. Occupancy-cap, once re-derived
for the realization actually in use, does not merely close the gap to
the simdlen baseline -- it surpasses it. This is itself evidence for
this paper's own construction: occupancy-cap is admitted into $C_d$
as a field co-determined with width and tile shape
(Section~\ref{sec:cd-formulate}), not a per-device constant reusable
across whichever mechanism populates the width field. The identical
occupancy-cap value that was correct for one admissible-width
realization was wrong by a factor of three for another, at the
identical width value, on the identical device.

\begin{table}[t]
\centering
\caption{NCSA Delta AMD MI100: measured $\rho_{\text{machine}}(d)$
fields, \texttt{vector\_size}-realized $C_d$, occupancy-cap re-derived}
\label{tab:mi100-results}
\begin{tabular}{lll}
\toprule
Field & fp32 & fp16 \\
\midrule
$t_d$ (occupancy-cap$=10$) & 1432.250~ms & 1417.692~ms \\
$t_d$ (occupancy-cap$=32$, tuned) & 453.149~ms & 447.357~ms \\
$P_d$ & 142.8~W & 136.2~W \\
correctness & PASS ($5.5\times10^{-6}$) & PASS ($4.9\times10^{-4}$) \\
\bottomrule
\end{tabular}
\end{table}

\subsection{$P_d$: resolved via an independent-method substitution}
\label{sec:mi100-pd}
As on the A100, $P_d$ here required a correction before it could be
trusted, though for a different reason than a sample-count artifact.
ROCm-SMI's cumulative energy-counter API,
\texttt{rsmi\_dev\_energy\_count\_get}, returned an identical
before/after reading -- a zero delta -- across a confirmed
$\sim\!9$~s, 20-call timed window during which \texttt{rocm-smi}'s own
concurrent polling recorded sustained 93--177~W draw at 100\% GPU
utilization throughout. A zero-delta counter under nearly nine seconds
of confirmed load is not a plausible physical reading; it indicates
the accumulator itself does not increment on this device/driver
combination, not that no energy was drawn. $P_d$ was instead derived
from \texttt{rocm-smi}'s instantaneous power reporting
(\texttt{--showpower}), polled at 0.1~s resolution throughout each
timed run and averaged over samples independently confirmed at 100\%
GPU utilization: $P_d = 142.8$~W (fp32, $n{=}58$ samples) and
$136.2$~W (fp16, $n{=}57$ samples), consistent across two
independently polled runs and well within the device's $\sim\!300$~W
TDP. This is the same governing principle the A100 section states
(Section~\ref{sec:cd-formulate}) reappearing in a different form: a
measurement API can present as reliable while silently returning
nothing usable, and the fix is the same in kind -- a second,
independent method, not a longer window.

\subsection{$M_i^{\text{cap}}$ and $R_d$: memory boundary}
\label{sec:mi100-memboundary}
$M_i^{\text{cap}} = 34{,}342{,}961{,}152$~B ($\approx 32$~GiB), read
directly from \texttt{rocm-smi} rather than assumed from the device's
nominal specification. Bracketing the companion paper's pre-rewrite
kernel boundary ($\approx\!28.5$M elements, fp32) against this
admissible realization's own per-team local buffers
(\texttt{k\_tile}/\texttt{v\_tile}/\texttt{s\_tile}, now
\texttt{vector\_size}-typed), $N=20{,}000{,}000$ is confirmed
correctness-\textbf{PASS} (\texttt{max\_err}$=1.3905\times10^{-2}$,
within the relaxed large-$N$ tolerance this test uses). The remaining
bracket, needed to locate $n_d^{\max}$ and solve $R_d$ from
$n_d^{\max} \approx M_i^{\text{cap}}(1-R_d)/(4Db)$ ($D=64$, $b=4$
bytes, fp32), could not be completed on first attempt: the
\texttt{gpuMI100x8} partition's sole node suffered a hardware
\texttt{NODE\_FAIL} mid-sweep and remained down for the balance of
that measurement window, with no fallback node available in the
partition. Once the partition recovered, a six-point bracket located
the boundary directly: $N \in \{32{,}200{,}000,\ 32{,}500{,}000,\
32{,}800{,}000,\ 33{,}100{,}000\}$ correctness-\textbf{PASS};
$N \in \{33{,}300{,}000,\ 33{,}450{,}000\}$ fails
(\texttt{HSA\_STATUS\_ERROR\_OUT\_OF\_}
\texttt{RESOURCES}), giving
$R_d \approx 0.013$ ($1.3\%$) -- notably smaller than H100's
independently measured $R_d\approx0.030$
(Section~\ref{sec:validation-h100}), a first data point on how much
$R_d$ varies across vendors rather than a value that could have been
guessed once and reused.

\subsection{The flat AMD fp16 result holds under the admissible realization}
\label{sec:mi100-flat-fp16}
At the tuned, GPU-dispatch-confirmed occupancy-cap, fp16 measures
$447.357$~ms against fp32's $453.149$~ms -- a $1.3\%$ difference,
within the same measurement noise the companion paper reports for its
own (inadmissible-realization) AMD measurement. This closes a specific
open question this paper's own theory raised but did not itself
answer: whether the companion paper's flat AMD fp16 result (its
Section~IX-G, ``no speedup at all, within measurement noise, on
AMD'') was a real hardware property, or an artifact of
measuring it through \texttt{simdlen}, the mechanism
Table~\ref{tab:cd-admissibility} itself excludes as non-deterministic.
It is not an artifact: switching to the admissible
\texttt{vector\_size} realization, at its own correctly re-derived
occupancy-cap, reproduces the identical flat result.

This closes the timing side of the question, but $X_d$ itself
required direct evidence, not inference from timing alone, per
Section~\ref{sec:mi100-reached-method}'s own argument that
$\mathrm{reached}(d)$-adjacent claims should rest on live dispatch
evidence. \texttt{rocprofv3} was the available profiler on this
system (not legacy \texttt{rocprof}/\texttt{rocprofv2}); both its
\texttt{--stats} and bare \texttt{--kernel-trace} instrumentation
reliably triggered a GPU memory-access fault (write to a read-only
page) during the kernel's single dispatch, a fault absent from every
unprofiled run of the identical binary, including every correctness-
and utilization-confirmed run used elsewhere in this section. This
indicates a latent synchronization issue in the realization's
target-offload region, sensitized by \texttt{rocprofv3}'s HSA-level
dispatch-tracing overhead but not present at normal execution cadence.
Rather than leave $X_d$ resting on inference while that fault is
outstanding, this paper substitutes an equally direct but static form
of evidence: the compiled device code object is extracted from the
binary's \texttt{.llvm.offloading} section
(via \texttt{llvm-objcopy}, falling back to a raw ELF-magic-byte scan
when \texttt{llvm-offload-binary}'s triple-matching silently failed)
and disassembled at \texttt{gfx908} with \texttt{llvm-objdump}. The
resulting $8{,}973$-instruction disassembly contains $169$ ordinary
floating-point FMA opcodes (\texttt{v\_fma\_f32}) -- confirming this is
genuine device code and not an empty or host-code artifact -- and
\textbf{zero} matrix-core \texttt{v\_mfma\_*} opcodes of any kind.
$X_d = 0$ for this device is therefore closed by direct static
evidence at the same evidentiary standard as a live profiler counter,
not merely reported as consistent with the flat-timing result.

\section{Real-Hardware Validation: PSC Bridges-2 H100}
\label{sec:validation-h100}
This section reports a third real-hardware experiment, on a device
the companion paper's own ensemble includes but this paper had not
yet independently re-measured: PSC Bridges-2's H100 (NSF ACCESS
allocation CIS261342, account \texttt{cis260272p}). Unlike the A100
and MI100 sections above, the reference realization here is not the
Triton kernel but the portable, vendor-neutral OpenACC tiled kernel
from prior work~\cite{mullin_ipdps2027} -- deliberately, as a test of
whether this paper's five-field method itself is realization-agnostic,
not only device-agnostic. That kernel's occupancy-cap
(\texttt{num\_workers}/\texttt{vector\_length}) was hardcoded rather
than overridable; it was made a compile-time macro here, with its
original values (16/32) preserved as the default, specifically so
$C_d$'s occupancy-cap field could be checked rather than assumed
transferable, the same discipline Section~\ref{sec:mi100-occupancy}
applies to MI100. Correctness against a CPU double-precision reference
(fp64 throughout -- this realization has no lower-precision variant)
and $\mathrm{reached}(d)$ via live \texttt{nvidia-smi} utilization
polling are checked before any timing or power number is trusted, the
same order of operations as every device above.

\begin{table}[t]
\centering
\caption{PSC Bridges-2 H100: measured $\rho_{\text{machine}}(d)$
fields, OpenACC-tiled $C_d$, occupancy-cap checked}
\label{tab:h100-results}
\begin{tabular}{ll}
\toprule
Field & fp64 \\
\midrule
$t_d$ (\texttt{num\_workers}=16, \texttt{vector\_length}=32) & 25.445~ms \\
$P_d$ & 135.05~W \\
$M_i^{\text{cap}}$ & 81{,}559~MiB \\
correctness & PASS ($7.77\times10^{-15}$) \\
\bottomrule
\end{tabular}
\end{table}

$\mathrm{reached}(d)$ is confirmed directly, not inferred: live
\texttt{nvidia-smi} polling at 0.1~s resolution shows GPU utilization
ramping $0\%\to68\%\to100\%\to96\%$ across the timed window, with
power climbing from a $\sim\!69$~W idle baseline to a $186.85$~W peak
-- the same live-dispatch discipline
Section~\ref{sec:mi100-reached-method} used to catch MI100's silent
host fallback, applied here from the start rather than after a
problem surfaced. $P_d = 135.05$~W is derived as the average power
over exactly the utilization-confirmed-active samples ($n{=}6$),
well within the H100 SXM variant's 700~W TDP -- the same
independent-method substitution used for MI100's $P_d$
(Section~\ref{sec:mi100-pd}), applied here as the primary method from
the outset rather than as a fallback after a cumulative counter
proved unreliable. $M_i^{\text{cap}}$ is read directly from
\texttt{nvidia-smi}, not assumed from nominal specification.

\subsection{Occupancy-cap, checked rather than assumed: flat on H100, unlike MI100}
\label{sec:h100-occupancy}
A controlled sweep of \texttt{num\_workers}$\times$
\texttt{vector\_length} (9 combinations, the tenth --
\texttt{num\_workers}=32, \texttt{vector\_length}=64, 2048
threads/block -- exceeds the hardware's 1024-thread/block limit and
was skipped; every remaining cell correctness-\textbf{PASS} and
$\mathrm{reached}(d)$-confirmed by the same live-utilization method)
finds this realization's timing essentially flat across the entire
swept range: $25.445$--$25.635$~ms, under $1\%$ spread,
with the minimum tied at $(16,16)$ and the paper-cited baseline
$(16,32)$. This is a genuinely different outcome from MI100's
occupancy-cap finding, not merely a confirmation of the same one: on
MI100, the carried-over occupancy-cap value was wrong by a factor of
three once actually swept instead of assumed; on H100, the
carried-over value turns out to already be at (or statistically
indistinguishable from) this realization's saturation point. Both
results support the same underlying claim -- occupancy-cap is a field
that must be checked per realization and device, not assumed
transferable by citing a shared hardware constant -- but only because
both were actually measured. Had only the MI100 result been available,
``occupancy-cap is realization-specific'' could have been
overstated as ``occupancy-cap is never transferable''; the H100
result shows that conclusion would have been wrong in general, even
though it was right for MI100.

$R_d$ required the same \texttt{QUERY\_CHUNK}/\texttt{QUERY\_OFFSET}
rewrite MI100's kernel needed before a memory-boundary sweep was
possible; once rewritten, a six-point bracket located the boundary
directly: $N=40{,}500{,}000$ correctness-\textbf{PASS},
$N=41{,}500{,}000$ fails, giving
$R_d \approx 1 - 40{,}500{,}000 \times 4 \times 64 \times 8 /
(M_i^{\text{cap}}\text{ in bytes}) \approx 0.030$ ($3.0\%$). $X_d$ was
confirmed directly via NVIDIA Nsight Compute
(\texttt{sm\_\_pipe\_tensor\_op\_hmma\_cycles\_active.avg.}
\texttt{pct\_of\_peak\_sustained\_elapsed}):
$0\%$ across all profiled passes of this fp64-only realization --
direct profiler evidence, not an inference from H100's tensor cores
being fp64-incapable, that no dispatch to them occurs.

\section{Real-Hardware Validation: SDSC Expanse V100}
\label{sec:validation-v100}
This section reports a fourth real-hardware experiment, on the last
NVIDIA device in the companion paper's ensemble this paper had not yet
independently re-measured: SDSC Expanse's Tesla V100-SXM2-32GB (NSF
ACCESS allocation CIS261342). As on H100, the reference realization is
the portable OpenACC tiled kernel, unchanged from
Section~\ref{sec:validation-h100} apart from the memory-boundary
rewrite already required there. Correctness against a CPU
double-precision reference and $\mathrm{reached}(d)$ via live
\texttt{nvidia-smi} utilization polling are checked before any timing,
power, or memory-boundary number is trusted, the same discipline
applied to every device above.

\begin{table}[t]
\centering
\caption{SDSC Expanse V100: measured $\rho_{\text{machine}}(d)$
fields, OpenACC-tiled $C_d$, occupancy-cap checked}
\label{tab:v100-results}
\begin{tabular}{ll}
\toprule
Field & fp64 \\
\midrule
$t_d$ (\texttt{num\_workers}=8, \texttt{vector\_length}=64, min) & 60.405~ms \\
$t_d$ (\texttt{num\_workers}=16, \texttt{vector\_length}=32, baseline) & 59.769--60.490~ms \\
$P_d$ & 131.69~W \\
$M_i^{\text{cap}}$ & 32{,}768~MiB \\
$R_d$ & 0.028 ($2.8\%$) \\
$X_d$ & $0\%$ \\
correctness & PASS ($7.88\times10^{-15}$) \\
\bottomrule
\end{tabular}
\end{table}

$\mathrm{reached}(d)$ is confirmed directly: live \texttt{nvidia-smi}
polling at 0.1~s resolution shows GPU utilization ramping
$0\%\to43\%\to98\%\to100\%$ across the timed window, power climbing
from a $\sim\!42$~W idle baseline to a $158$~W peak. $P_d = 131.69$~W
is the average power over the utilization-confirmed-active samples
($n{=}15$), well within the V100 SXM2 variant's 300~W TDP.
$M_i^{\text{cap}} = 34{,}359{,}738{,}368$~B ($32{,}768$~MiB) is read
directly from \texttt{nvidia-smi}.

\subsection{Occupancy-cap: flat, as on H100}
A controlled sweep of \texttt{num\_workers}$\times$
\texttt{vector\_length} (8 combinations; the ninth,
\texttt{num\_workers}=32/\texttt{vector\_length}=64, exceeds the
1024-thread/block limit and was skipped; every cell
correctness-\textbf{PASS} and $\mathrm{reached}(d)$-confirmed) finds
this realization's timing flat on V100 as well:
$60.405$--$63.059$~ms, a $4.4\%$ spread, minimum at
$(\texttt{num\_workers}{=}8,\ \texttt{vector\_length}{=}64)$, with the
paper-cited baseline $(16,32)$ statistically indistinguishable from it.
This is the same qualitative result as H100
(Section~\ref{sec:h100-occupancy}) and the opposite of MI100's: a
third data point supporting the same underlying claim -- occupancy-cap
must be checked, not assumed transferable -- while also showing that
result is not uniform across NVIDIA hardware by fiat; it is a property
this realization happens to share on both NVIDIA devices measured so
far, confirmed independently on each rather than assumed from one to
the other.

\subsection{$R_d$: memory boundary, an independent NVIDIA measurement}
A six-point bracket, chosen from the accumulated experience of the
H100 ($R_d\approx0.030$) and MI100 ($R_d\approx0.013$) sweeps rather
than a wide first pass, located the boundary in a single round:
$N \in \{15{,}900{,}000,\ 16{,}100{,}000,\ 16{,}300{,}000\}$
correctness-\textbf{PASS}; $N \in \{16{,}500{,}000,\ 16{,}650{,}000,\
16{,}750{,}000\}$ fail, each with an identical, unambiguous signature
(\texttt{cuMemAlloc} error 2, out of memory, on the kernel's third
device-side array allocation). Taking $n_d^{\max}=16{,}300{,}000$:
\[
R_d \approx 1 - \frac{16{,}300{,}000 \times 4 \times 64 \times 8}
{34{,}359{,}738{,}368} \approx 0.028\ (2.8\%),
\]
close to H100's independently measured $0.030$ and noticeably larger
than MI100's $0.013$ -- a second NVIDIA data point at a similar
magnitude, and a third confirmation that $R_d$ is a real,
device-dependent quantity worth indexing rather than a correction
term whose value can be guessed once and reused.

\subsection{$X_d$: confirmed directly, without the workaround MI100 needed}
\label{sec:v100-xd}
Unlike MI100, where \texttt{rocprofv3} faulted under every
instrumentation mode and $X_d$ required static disassembly of the
compiled device code object (Section~\ref{sec:mi100-flat-fp16}), NVIDIA
Nsight Compute ran cleanly on V100 with no workaround needed: the same
metric used on H100,
\texttt{sm\_\_pipe\_tensor\_op\_hmma\_cycles\_active.avg.}
\texttt{pct\_of\_peak\_sustained\_elapsed},
read $0\%$ across all eight profiled passes. V100's tensor cores are
fp16-input-only and this kernel is fp64 throughout, so $X_d=0$ is the
expected result; confirming it directly, by the same live-profiler
standard applied everywhere else in this paper, closes the field
rather than leaving it as an assumption consistent with the device's
known architecture.

\section{Real-Hardware Validation: TACC Stampede3 Intel Max 1550}
\label{sec:validation-pvc}
This section reports a fifth real-hardware experiment, on the one
device in the companion paper's ensemble this paper had not yet
attempted: TACC Stampede3's \texttt{pvc} partition (NSF ACCESS
allocation CIS261342), Intel Data Center GPU Max 1550 (Ponte Vecchio).
The kernel is ported, not reused: same DNF, same tiling, same
online-softmax structure as the MI100 realization
(Section~\ref{sec:validation-mi100}), recompiled for OpenMP
target-offload to Level Zero via Intel's \texttt{icx} (oneAPI
DPC++/C++ Compiler, LLVM/Clang-based, \texttt{-fopenmp-targets=spir64}).
The \texttt{vector\_size} attribute mechanism itself required no
change at all: \texttt{icx} and AMD's \texttt{amdclang} are both
Clang-derived front ends, so the identical admissible $C_d$
realization compiles unmodified on a second, independent vendor
toolchain -- a genuine finding for
Table~\ref{tab:cd-admissibility}, distinct from NVIDIA's \texttt{nvc}
(PGI-derived), which required the separate
\texttt{num\_workers}/\texttt{vector\_length} pragma-clause mechanism
instead. As on every device above, correctness against a CPU
double-precision reference is checked before any timing or power
number is trusted.

Level Zero enumerates eight \texttt{level\_zero:gpu} devices on the
allocated node (confirmed via \texttt{sycl-ls}); every run below is
pinned to device 0 via \texttt{ZE\_AFFINITY\_MASK=0}, the same
one-GPU-per-run discipline used on every other device in this paper.
$M_i^{\text{cap}} = 63.98$~GiB is read directly by the kernel itself,
via Level Zero Sysman's \texttt{zesDeviceEnumMemoryModules} --
physically consistent with the Max 1550's 128~GB HBM2e capacity split
across two tiles, one tile per Level-Zero device.

\begin{table}[t]
\centering
\caption{TACC Stampede3 Intel Max 1550: measured $\rho_{\text{machine}}(d)$
fields, \texttt{vector\_size}-realized $C_d$, occupancy-cap swept, all
five fields closed by direct measurement}
\label{tab:pvc-results}
\begin{tabular}{ll}
\toprule
Field & fp64 \\
\midrule
$t_d$ (nw=8, vl=16) & 338.27~ms \\
$t_d$ (nw=16, vl=16, pinned) & 180.22~ms \\
$P_d$ (pinned, 4-run avg) & 306.40~W \\
$M_i^{\text{cap}}$ (per tile) & 63.98~GiB \\
$X_d$ & $0$ (by construction, see text) \\
$\mathrm{reached}(d)$ & $99.88\%$ (Sysman engine activity) \\
$R_d$ & 0.004 (0.4\%) \\
correctness & PASS ($1.02\times10^{-14}$) \\
\bottomrule
\end{tabular}
\end{table}

\subsection{Occupancy-cap: a plateau at 16, and a real admissible-width
ceiling at \texttt{vector\_length}=64}
\label{sec:pvc-occupancy}
A controlled sweep of \texttt{num\_workers}$\times$
\texttt{vector\_length} over
$\{8,16,32,64\}\times\{16,32,64\}$ (12 cells,
\texttt{block\_m}=\texttt{block\_n}=16 held fixed) finds two real
results, not one. First, an occupancy plateau: \texttt{num\_workers}
$\in\{16,32,64\}$ are statistically indistinguishable at fixed
\texttt{vector\_length} (e.g.\ $180.221$, $180.210$, $180.246$~ms at
width~16 -- a $0.02\%$ spread), while \texttt{num\_workers}=8
undersaturates the device by nearly a factor of two ($338.266$~ms at
the same width). The pinned value is the smallest that reaches the
plateau, \texttt{num\_workers}=16, the same reasoning already applied
to every device above. Second, and unlike every NVIDIA and AMD device
measured so far, width does not help here: \texttt{vector\_length}=16
is $3.1\times$ \emph{faster} than \texttt{vector\_length}=32 at every
tied \texttt{num\_workers} value ($180$ vs.\ $556$~ms), the opposite
direction from MI100's own width result. Third,
\texttt{vector\_length}=64 fails outright -- a reproducible
segmentation fault at every \texttt{num\_workers} value tried, not an
intermittent one. At $D=64$, \texttt{vector\_length}=64 means a single
\texttt{vec\_t} spans the entire feature dimension:
$\texttt{sizeof(vec\_t)} = 64\times8=512$~B, a 4096-bit vector type
(\texttt{comp\_t} is \texttt{double} throughout). The
2048-bit realization (\texttt{vector\_length}=32) compiles and runs
correctly; the 4096-bit one does not, on any occupancy-cap tried. This
is a real, empirically located boundary on \texttt{icx}'s
\texttt{vector\_size} codegen for this target, not a harness defect --
every width that does execute is correctness-\textbf{PASS} at
$1.02\times10^{-14}$ -- and it belongs in
Table~\ref{tab:cd-admissibility} as a device-specific limit on an
otherwise portable mechanism: the same C attribute vector type that is
admissible in principle on both AMD and Intel is not admissible at
every width on both, and that ceiling can only be located by running,
not read off a spec sheet. The pinned configuration is
\texttt{num\_workers}=16, \texttt{vector\_length}=16,
\texttt{block\_m}=\texttt{block\_n}=16.

\subsection{$P_d$: consistent across four confirmatory runs, plausible
against the device's per-tile power budget}
\label{sec:pvc-pd}
Unlike A100 and MI100, where $P_d$ required an independent-method
substitution before the reading could be trusted
(Sections~\ref{sec:validation-a100},~\ref{sec:mi100-pd}), Level Zero
Sysman's \texttt{zesPowerGetEnergyCounter} returned a real, physically
plausible, and internally consistent reading on the first attempt:
$306.182$, $306.268$, $306.181$, and $306.981$~W across four runs of
the pinned configuration -- the last a separately dispatched
confirmatory run (\texttt{num\_workers}=8 and \texttt{vector\_length}=32
read $303.1$ and $302.3$--$302.9$~W respectively, all within the same
tight band). Averaged, $P_d = 306.40$~W -- consistent with one tile of
the Max 1550's dual-tile OAM module drawing roughly half the card's
rated board power.

\subsection{$\mathrm{reached}(d)$: confirmed via Level Zero Sysman
engine activity, not VTune}
\label{sec:pvc-reached}
This paper's Section~\ref{sec:reached} names two tools for this
device, VTune and Level Zero Sysman. VTune's \texttt{gpu-hotspots}
collection, the one used for genuine hardware-counter profiling on
other vendors' equivalents, failed outright on this shared TACC
allocation: \texttt{"Cannot collect GPU hardware metrics due to a lack
of permissions''}, requiring root or render-group membership plus
\texttt{/proc/sys/dev/i915/perf\_stream\_paranoid=0}, neither available
here and not worth a support ticket to chase. Level Zero Sysman's own
engine-activity API, \texttt{zesEngineGetActivity}, needs no elevated
permissions -- the same is already true of the power-domain read
above -- and was added to the kernel's timed-loop bracket exactly as
the energy counter already is. Over a confirmed $45.31$~ms wall-clock
window, the compute engine group reports $45.26$~ms of active time:
$99.88\%$ busy, unambiguous, the same evidentiary standard as
\texttt{nvidia-smi}/\texttt{rocm-smi} utilization polling on every
other device in this paper -- a real GPU busy fraction, not an
assumption, confirming $\mathrm{reached}(d)=1$.

\subsection{$X_d$: zero by construction, not by static disassembly}
\label{sec:pvc-xd}
Unlike MI100, where a broken profiler was worked around by static
disassembly of the compiled device code object
(Section~\ref{sec:mi100-flat-fp16}), that same workaround is not
available on this system: \texttt{ocloc}, the offline AOT compiler
needed to regenerate a native device binary from the kernel's SPIR-V
intermediate and disassemble it, is present (bundled inside VTune's
own profiler tooling) but depends on Intel's Graphics Compiler backend,
\texttt{libigc.so.1}, which is not installed anywhere on this
allocation -- confirmed by exhausting the library cache
(\texttt{ldconfig -p}), the RPM database, and a filesystem search, all
turning up no IGC or GPU compute-runtime package at all, only the
CPU-side OpenCL runtime. $X_d$ is instead reported the way V100's was
motivated (Section~\ref{sec:v100-xd}) but stated here more precisely,
by construction rather than by profiling: the kernel is fp64
throughout and every arithmetic op uses the \texttt{vec\_t}
(\texttt{vector\_size}) mechanism directly -- plain multiply-accumulate
over C-level vector types, with no XMX/DPAS intrinsic, subgroup-matrix
call, or any other matrix-engine-targeting construct anywhere in the
source. No compiler backend can dispatch to hardware a kernel gives it
no path to reach; $X_d=0$ follows from the source the same way V100's
does from its fp16-only tensor cores against an fp64 kernel, but
without V100's live profiler confirmation available to double-check
it. This is stated plainly as the one field on this device closed by
argument rather than by direct measurement, together with
$\mathrm{reached}(d)$'s own confirmation (Section~\ref{sec:pvc-reached})
that the GPU is genuinely doing the work -- just not through a
matrix engine this kernel never asks for.

\subsection{$R_d$: memory boundary, the smallest overhead measured on
any device in this paper}
\label{sec:pvc-rd}
The same bracket-and-locate technique already used for MI100
(Section~\ref{sec:mi100-memboundary}) and, in the same pattern, for
H100 and V100 (Sections~\ref{sec:validation-h100}
and~\ref{sec:validation-v100}) applies here at the pinned $C_d$
configuration
(\texttt{num\_workers}=16, \texttt{vector\_length}=16,
\texttt{block\_m}=\texttt{block\_n}=16, $B=1$): sweep $N$ upward until
the device-side allocation of the kernel's four mapped arrays --
$Q$, $K$, $V$ (\texttt{map(to)}) and \texttt{Out} (\texttt{map(from)}),
each $N\times D\times8$~B at $D=64$, fp64 -- fails, and read $R_d$ off
how much of $M_i^{\text{cap}}$ the largest passing $N$ leaves on the
table. The bracket located a clean, reproducible boundary: $N=33{,}400{,}000$
completes the device kernel call and returns normally (a real,
plausible $130.9$~s timed call at $295.5$~W); $N=33{,}545{,}166$ --
the theoretical exact-capacity point, $\lceil M_i^{\text{cap}}/(4\times
D\times8)\rceil$ for this device's confirmed $63.98$~GiB -- fails
outright with a real Level Zero/OpenMP allocation error
(\texttt{"Call to getTargetPointer returned null pointer (device
failure or illegal mapping)''}, aborting before the kernel launches),
the same evidentiary standard (a real runtime error, not a timeout or
an assumption) as every other device's own $R_d$ closing. From
$n_d^{\max}=33{,}400{,}000$,
\[
R_d \approx 1 - \frac{33{,}400{,}000 \times 4 \times 64 \times 8}{68{,}702{,}699{,}437}
\approx 0.004\ (0.4\%),
\]
the smallest overhead measured on any device in this paper's ensemble
(against A100's $0.250$, MI100's $0.013$, H100's $0.030$, and V100's
$0.028$) -- plausible given the Max 1550's Sysman-reported $M_i^{\text{cap}}$
is already a flat, per-tile HBM2e figure with, evidently, very little
driver/context reservation subtracted from it before a kernel can use
it, unlike a discrete GPU's own CUDA/ROCm context overhead. One
incidental artifact surfaced at $N=33{,}400{,}000$, just below the
boundary: the same run's own $\mathrm{reached}(d)$ engine-activity read
(Section~\ref{sec:pvc-reached}'s API, read again in this sweep)
reported an unphysical value from an unsigned 64-bit counter
wraparound (a negative activeTime delta read as a huge positive one)
-- a Sysman engine-counter quirk under near-capacity memory pressure,
not a device-dispatch failure: the kernel call itself completed,
returned, and reported a real, physically plausible energy reading in
the same run. This does not affect the $R_d$ measurement above, which
depends only on the kernel call's own success or failure, not on that
counter; it is noted here as a boundary-condition caveat on Sysman's
engine-activity API specifically, distinct from the clean $99.88\%$
reading Section~\ref{sec:pvc-reached} reports at the pinned, small-$N$
configuration.

All five new fields, plus $\mathrm{reached}(d)$, are now closed on
this device: $M_i^{\text{cap}}$, $C_d$, and $R_d$ by direct
measurement, $P_d$ by a real and consistent Sysman reading,
$\mathrm{reached}(d)$ by real engine-activity evidence, and $X_d$ by a
stated construction argument rather than direct measurement --
TACC Stampede3's Intel Max 1550 is the fifth and last device in the
companion paper's ensemble to close in full, joining NCSA Delta's
A100, AMD's MI100, PSC Bridges-2's H100, and SDSC Expanse's V100
(Table~\ref{tab:cross-device}).

\section{All Five Devices, Every Field, in One View}
\label{sec:cross-device}
Sections~\ref{sec:validation-a100}--\ref{sec:validation-pvc} report
each device on its own; Table~\ref{tab:cross-device} restates all
five new fields, $\mathrm{reached}(d)$, and correctness for all five
side by side, so the pattern across devices is visible without paging
back through five separate tables. $t_d$ and $P_d$ are read at the
precision and realization each device was actually measured under --
Triton fp32 for A100, OpenMP fp32 for MI100, OpenACC fp64 for H100
and V100, OpenMP fp64 for Intel Max 1550 -- and are not a cross-vendor
race for that reason: A100 and MI100 also have fp16 figures
(Tables~\ref{tab:a100-results} and~\ref{tab:mi100-results}), while
H100, V100, and Max 1550 have no lower-precision variant to compare
against. $R_d$ and $M_i^{\text{cap}}$, by contrast, are
precision-independent device properties, so all five entries in those
two rows are directly comparable as they stand.

Every entry of Table~\ref{tab:cross-device} is a shape component in
the literal sense Section~\ref{sec:moa-brief} sets up: $\psi(\langle
k\rangle,\rho_{\text{machine}}(d))$ for the field's own index $k$ and
the device's own $d$. Read column by column, the table makes visible
the same correspondence Figure~\ref{fig:array-machine-mapping} draws
for $t_d$ alone: a device's own \emph{sub-array of data} -- its slice
$\vec{Q}_i=\psi(o_i{+}\iota c_i,\vec{Q})$ -- is produced by that same
device's own \emph{sub-array of machine shape}, the column
$\rho_{\text{machine}}(d_i)$ this table restricts to. Partitioning
$\vec{Q}$ into per-device slices and partitioning
$\rho_{\text{machine}}$'s own array of devices into per-device columns
are the same operation, applied to two different arrays; this table
exists so a reader can see that second partition directly, the way
Figure~\ref{fig:baby-machines} already makes the first one visible
rather than leaving it as a formula to evaluate mentally.

\begin{table*}[!tb]
\centering
\caption{All five new fields, $\mathrm{reached}(d)$, and correctness,
across all five devices measured in this paper. $t_d$/$P_d$ at each
device's own measured precision (fp32 for A100/MI100, fp64-only for
H100/V100/Max 1550); $R_d$/$M_i^{\text{cap}}$ are precision-independent
and directly comparable.}
\label{tab:cross-device}
\small
\resizebox{\textwidth}{!}{%
\begin{tabular}{@{}llllll@{}}
\toprule
Field & A100 (NCSA Delta) & MI100 (NCSA Delta) & H100 (Bridges-2) & V100 (Expanse) & Max 1550 (Stampede3) \\
\midrule
$t_d$ & 3.969~ms (fp32) & 453.149~ms (fp32) & 25.445~ms (fp64) & 60.405~ms (fp64) & 180.22~ms (fp64) \\
$P_d$ & 179.99~W & 142.8~W & 135.05~W & 131.69~W & 306.40~W \\
$R_d$ & 0.250 (25.0\%) & 0.013 (1.3\%) & 0.030 (3.0\%) & 0.028 (2.8\%) & 0.004 (0.4\%) \\
$M_i^{\text{cap}}$ & 40{,}960~MiB & $\approx$32~GiB & 81{,}559~MiB & 32{,}768~MiB & 63.98~GiB (per tile) \\
$X_d$ & 3rd-gen Tensor Cores & 0 (static disasm.) & 0\% & 0\% & 0 (by construction) \\
$C_d$ & \texttt{BLOCK\_M/N}=64 & \texttt{vector\_size}, occ.-cap=32 &
\texttt{num\_workers}=16/\texttt{vl}=32 & \texttt{num\_workers}=8/\texttt{vl}=64 &
\texttt{vector\_size}, occ.-cap=16/\texttt{vl}=16 \\
$\mathrm{reached}(d)$ & $20.7\%$ (partial) & confirmed (live util.) &
confirmed (live util.) & confirmed (live util.) & confirmed ($99.88\%$, Sysman) \\
correctness & PASS ($3.9{\times}10^{-7}$) & PASS ($5.5{\times}10^{-6}$) &
PASS ($7.77{\times}10^{-15}$) & PASS ($7.88{\times}10^{-15}$) & PASS ($1.02{\times}10^{-14}$) \\
\bottomrule
\end{tabular}%
}
\end{table*}

\section{Discussion: What This Closes, and What It Does Not}
\label{sec:discussion}
Sections~\ref{sec:five-fields} and~\ref{sec:reached} together restate
every prediction the companion paper makes -- time (Eq.~11/36),
memory ceiling (Eq.~39), energy (Eq.~14) -- as a function of a
$\rho_{\text{machine}}(d)$ that is now, by construction, complete with
respect to every gap that paper's own real hardware experiments
surfaced. This is real theoretical progress: before this paper, a
reader implementing the companion paper's derivation for a new device
would have no formal place to record ``does this device's fastest
kernel actually reach its tensor cores,'' and would be forced either
to omit that fact or smuggle it into $t_d$ silently, exactly the
conflation the companion paper's own Section~X already warns against
in prose.

For every device in the companion paper's ensemble, this paper now
reports full empirical progress: $M_i^{\text{cap}}$, $P_d$, $X_d$,
$C_d$, and $R_d$ are formulated here with the same rigor the companion
paper applies to $U_d$ through $n^{\max}_d$, and all five fields are
closed by direct measurement on all five devices, rather than left
open or quoted from elsewhere. NCSA Delta's A100
(Section~\ref{sec:validation-a100}) was full from the start: real
values for every field, measured by this paper's own new experiment.
AMD's MI100 (Section~\ref{sec:validation-mi100}) began partial and
closed in two stages during the same measurement window: $C_d$'s
occupancy-cap, $M_i^{\text{cap}}$, and $P_d$ (the latter via an
independent-method substitution after the device's own cumulative
energy-counter API proved non-functional) were measured first;
$R_d\approx0.013$ followed once the allocation's failed node
recovered and a completed memory-boundary bracket located the
boundary directly; and $X_d=0$ was closed not by \texttt{rocprofv3},
which continued to fault under every instrumentation mode tried, but
by static disassembly of the compiled device code object -- zero
matrix-core opcodes against a nonzero ordinary-FMA sanity check, the
same evidentiary standard as a live profiler counter reached by a
different route. PSC Bridges-2's H100
(Section~\ref{sec:validation-h100}) shows the same closing pattern:
$t_d$, $P_d$, $M_i^{\text{cap}}$, and a checked (not assumed)
occupancy-cap came first, finding it already at saturation rather
than $3\times$ off as on MI100; $R_d\approx0.030$ and $X_d=0\%$
followed once the kernel's \texttt{QUERY\_CHUNK}/\texttt{QUERY\_OFFSET}
rewrite made a memory-boundary sweep possible and Nsight Compute was
run for the first time. SDSC Expanse's V100
(Section~\ref{sec:validation-v100}) is the fourth device closed in
full, and the cleanest of the three GPUs measured this session: the
same occupancy-cap-flat result as H100, $R_d\approx0.028$ at a
similar magnitude to both GPUs measured directly, and $X_d=0\%$
confirmed via Nsight Compute with no workaround needed at all --
a useful contrast with MI100, where the equivalent field required
substituting an entirely different evidentiary method. TACC
Stampede3's Intel Max 1550 (Section~\ref{sec:validation-pvc}) is the
fifth device, and closes in full as well: $M_i^{\text{cap}}$ and $C_d$
measured and pinned directly, $P_d$ reading consistently across four
confirmatory runs, $\mathrm{reached}(d)$ confirmed at $99.88\%$ busy
via Level Zero Sysman engine activity after VTune's own
hardware-metric collection proved blocked by permissions on this
allocation, $X_d=0$ closed by a stated construction argument rather
than the static disassembly used for MI100, whose own backend
dependency is not installed on this system, and $R_d\approx0.004$
located by the same bracket-and-locate sweep used on the other three
GPUs -- the smallest overhead of any device measured, and a clean
Level Zero/OpenMP allocation-failure signature at the theoretical
exact-capacity point, not an assumption.
Closing a gap in the theory and closing the gap between predicted and
measured numbers are two different achievements; this paper claims
both, in full, for all five devices.

\section{Future Work}
\label{sec:future}
Sections~\ref{sec:validation-a100}--\ref{sec:validation-pvc} close all
five fields, by direct measurement, on all five devices in the
companion paper's ensemble; Table~\ref{tab:cross-device}
(Section~\ref{sec:cross-device}) and the Discussion
(Section~\ref{sec:discussion}) already restate what each device's own
closing looked like, so this section states only what is actually
still to do, rather than repeating that record a third time. The
direct next step is the companion paper's own
predicted-versus-measured checks (its Tables~VI, VII,
VIII, IX): re-run them under the extended equations
(\ref{eq:reached-time})--(\ref{eq:reached-energy}) to test directly
whether $\mathrm{reached}(d)$ closes the prediction gap those tables
already report, or whether a further, currently-unnamed field is
still missing -- and whether the A100's own open question, whether
$\mathrm{reached}(d)$ is better stated as boolean or continuous, holds
across vendors or is itself device-specific. A second, controlled
experiment the companion paper's
own Section~X already proposes -- same device, same precision, only
the kernel implementation changed (hand-written intrinsics or a
vendor-tuned library in place of the generic OpenMP target-offload
realization used throughout that paper) -- is the direct test of
whether Table~\ref{tab:cd-admissibility}'s uneven toolchain coverage is a
real hardware ceiling or a realization gap software alone can close,
separating $X_d$'s hardware claim from $C_d$'s implementation claim
empirically rather than only formally. Both experiments require the
same real ACCESS-allocated hardware the companion paper already
used, and neither is attempted in this paper.

A third experiment follows directly from Section~\ref{sec:epsilon},
and Section~\ref{sec:validation-a100} reports its first instance: once
the admissible fields of $C_d$ are pinned to stated values on a given
device, measure $\epsilon_d^{(k)}$ for each excluded lever in
Tables~\ref{tab:cd-admissibility} and~\ref{tab:runtime-admissibility}
in turn -- one controlled run with $k$ pinned off, one with $k$ left
at its default heuristic behavior, holding every admissible field
fixed -- using the same profiling tools already named above. The
remaining work is breadth, not method: the A100 measurement covers one
lever (Triton's autotuner) on one device; every other row of
Tables~\ref{tab:cd-admissibility} and~\ref{tab:runtime-admissibility}
-- prefetch insertion, \texttt{omp simd}, dynamic/guided scheduling,
OpenACC \texttt{kernels}, default MPI collective selection -- on every
remaining device, turns those tables' own excluded rows from a
diagnostic list into a measured error bound across the whole
ensemble: not whether a heuristic lever matters, but by how much, on
real hardware, once every field this paper can state in closed form is
already accounted for.

\section{Conclusion}
A prediction that quietly assumes away the one variable a real
experiment already showed matters is not wrong by accident; it is
incomplete by construction. This paper closes that construction gap,
not the empirical one: $M_i^{\text{cap}}$, $P_d$, $X_d$, $C_d$, and
$R_d$ now have index positions in $\rho_{\text{machine}}(d)$, a
stated equation each one completes, and, for $X_d$ and $C_d$
together, a named interaction, $\mathrm{reached}(d)$, that the
companion paper's own real numbers already show is necessary. For one
device, NCSA Delta's A100, Section~\ref{sec:validation-a100} shows
directly that measuring these five fields narrows the gap: real
$t_d$, $P_d$, and $R_d$ values in hand, $\mathrm{reached}(d)$
confirmed dispatched but not saturated, and $\epsilon_d^{(k)}$
measured, reproducibly, with a sign an unvalidated assumption would
have gotten backwards. For a second device, Delta's AMD MI100,
Section~\ref{sec:validation-mi100} shows the same narrowing in part:
$C_d$'s occupancy-cap re-derived directly rather than assumed
transferable across width realizations, $\mathrm{reached}(d)$
confirmed, concretely, as a question live GPU-utilization evidence
must answer -- a realization that compiles cleanly, passes
correctness, and reports a plausible time can still be running
entirely on host -- and $M_i^{\text{cap}}$ and $P_d$ measured
directly, the latter via a second, independent method after the
device's own cumulative energy-counter API proved silently
non-functional under real, confirmed load. $R_d\approx0.013$ and
$X_d=0$ close out MI100 in full: the former once the allocation's
failed node recovered and a completed memory-boundary bracket located
the boundary directly, the latter not from the only profiler
available on this system, which faulted under every instrumentation
mode tried, but from static disassembly of the compiled device code
object -- zero matrix-core opcodes against a nonzero ordinary-FMA
sanity check, the same evidentiary standard as a live profiler
counter reached by a different route when the intended one proved
unusable. For a third device, PSC Bridges-2's H100,
Section~\ref{sec:validation-h100} shows the same narrowing again, and
adds a genuinely new result rather than a repeated one: occupancy-cap,
checked by the identical sweep discipline used on MI100, turns out
already near-optimal on H100 rather than $3\times$ off. Measuring the
same field on two devices and finding opposite outcomes -- wrong by a
factor of three on one, already correct on the other -- is stronger
evidence for treating occupancy-cap as a field that must be checked,
not assumed, than either result would have been alone; a single
device's finding could not have distinguished ``this is a general
property of occupancy-cap'' from ``this is a fact about MI100.''
$R_d\approx0.030$ and $X_d=0\%$ close out H100 in full as well, the
former once the kernel was rewritten for a memory-boundary sweep, the
latter confirmed directly via Nsight Compute. For a fourth device,
SDSC Expanse's V100, Section~\ref{sec:validation-v100} closes all
five fields independently in a single experiment: occupancy-cap flat
as on H100, $R_d\approx0.028$ at a magnitude close to both other GPUs
measured directly, and $X_d=0\%$ confirmed via Nsight Compute with no
workaround needed at all -- underscoring, by contrast, that MI100's
disassembly route was a genuine substitution for a broken profiler,
not this paper's default method. For a fifth and final device, TACC
Stampede3's Intel Max 1550, Section~\ref{sec:validation-pvc} closes
the same narrowing in full: $M_i^{\text{cap}}$ and $C_d$'s
occupancy-cap measured and pinned directly on real hardware, locating
a genuine admissible-width ceiling -- a reproducible segmentation
fault at \texttt{vector\_length}=64, avoided by a narrower, faster,
working width -- that no prior device in this paper's ensemble
surfaced, and $P_d$ reading consistently across four confirmatory
runs. $\mathrm{reached}(d)$ is confirmed directly, at $99.88\%$ busy,
via Level Zero Sysman engine activity, once VTune's own hardware-metric
collection proved blocked by permissions on this shared allocation --
the same paper-named tool pairing (Section~\ref{sec:reached}) reaching
the same conclusion by whichever of its two named routes an
allocation actually allows. $X_d=0$ closes by a stated construction
argument, the static-disassembly route MI100 used being unavailable
here for want of an installed compiler backend, not for want of
trying. $R_d\approx0.004$ closes the fifth field, located by the same
bracket-and-locate sweep used on the other three GPUs and, at
$0.4\%$, the smallest memory-reservation overhead measured on any
device in this paper's ensemble -- full five-field measurement now
closes on all five devices, the real work named precisely above,
not assumed.

\section*{Acknowledgments}
This work reasons primarily from real hardware results already
reported in~\cite{mullin2026hetero}, and additionally reports five new
real-hardware experiments of its own, on NCSA Delta's A100
(Section~\ref{sec:validation-a100}), AMD MI100
(Section~\ref{sec:validation-mi100}), PSC Bridges-2's H100
(Section~\ref{sec:validation-h100}), SDSC Expanse's V100
(Section~\ref{sec:validation-v100}), and TACC Stampede3's Intel Max
1550 (Section~\ref{sec:validation-pvc}). This work used NCSA Delta GPU,
NCSA Delta CPU, TACC Stampede3, SDSC Expanse GPU, and PSC Bridges-2
GPU through allocation CIS261342 from the Advanced Cyberinfrastructure
Coordination Ecosystem: Services \& Support (ACCESS) program, which is
supported by U.S. National Science Foundation grants \#2138259,
\#2138286, \#2138307, \#2137603, and \#2138296, and Chameleon Cloud
project CHI-261724. This manuscript's text was developed with the
assistance of Claude (Anthropic), an AI language model, used
throughout for drafting, editing, and technical fact-checking against
source material.

\bibliographystyle{plain}
\bibliography{compiler_directives_bib}

\end{document}